\documentclass[%
 reprint,
superscriptaddress,
amsmath,amssymb,
aps,
pra,
]{revtex4-1}
\usepackage{caption}
\usepackage{subcaption}
\usepackage{graphicx}
\usepackage{xcolor}
\usepackage{graphicx}
\usepackage{dcolumn}
\usepackage{bm}

\begin{document}

\preprint{APS/123-QED}

\title{Engineering the optical vacuum: Arbitrary magnitude, sign, and order of dispersion in free space using space-time wave packets}

\author{Murat Yessenov}
\affiliation{CREOL, The College of Optics \& Photonics, University of Central Florida, Orlando, FL 32816, USA}
\author{Layton A. Hall}
\affiliation{CREOL, The College of Optics \& Photonics, University of Central Florida, Orlando, FL 32816, USA}
\author{Ayman F. Abouraddy}
\affiliation{CREOL, The College of Optics \& Photonics, University of Central Florida, Orlando, FL 32816, USA}
\email{*raddy@creol.ucf.edu}


\begin{abstract}
Spatial structuring of an optical pulse can lead in some cases upon free propagation to changes in its temporal profile. For example, introducing conventional angular dispersion into the field results in the pulse encountering group-velocity dispersion in free space. However, only limited control is accessible via this strategy. Here we show that precise and versatile control can be exercised in free space over the dispersion profile of so-called `space-time' wave packets: a class of pulsed beams undergirded by non-differentiable angular dispersion. This abstract mathematical feature allows us to tune the magnitude and sign of the different dispersion orders without introducing lossess, thereby realizing arbitrary dispersion profiles, and achieving dispersion values unattainable in optical materials away from resonance. Unlike optical materials and photonic structures in which the values of the different dispersion orders are not independent of each other, these orders are addressable separately using our strategy. These results demonstrate the versatility of space-time wave packets as a platform for structured light and points towards their utility in nonlinear and quantum optics.   
\end{abstract}

\maketitle

\section{Introduction}

The optical vacuum is non-dispersive. Unlike laser-pulse broadening in optical materials stemming from chromatic dispersion, propagation in free space does \textit{not} induce group-velocity dispersion (GVD) in plane-wave pulses \cite{SalehBook07}. However, \textit{spatial structuring} of the transverse field profile can lead to a variety of temporal changes in a pulse \cite{Akturk05OE,Dorrer19IEEE}. For example, the group velocity can be slightly reduced in free space by imposing particular spatial beam profiles on the pulsed field \cite{Giovannini15Science,Alfano16OC,Bouchard16Optica,Lyons18Optica} (although this is accompanied by GVD \cite{Zapata08JOSAA,Saari17OC}), and space-time coupling can lead to undesirable pulse deformation or reshaping, especially upon focusing short pulses \cite{Zhu05OE,SalehBook07}.

We are interested here in spatially structuring optical pulses to produce controllable GVD encountered by the freely propagating field in free space. Achieving this goal can have profound implications for nonlinear and quantum optics, where dispersion plays a crucial role in determining the outcome of multi-wavelength optical interactions with matter \cite{DiTrapani98PRL,Porras05OL,Faccio07OE,Malaguti08OL,Malaguti09PRA,Katamadze15PRA,Spasibko16OL,Cupita20OL}. One general strategy for spatially structuring the field is through introducing angular dispersion, which is a ubiquitous effect in optics whereby interaction with diffractive or dispersive devices such as gratings or prisms leads to a wavelength-dependent propagation angle \cite{Torres10AOP}. In general, angular dispersion tilts the pulse front with respect to the phase front \cite{Bor85OC,Hebling96OQE,Fulop10Review} and introduces effective GVD in free space \cite{Martinez84JOSAA,Porras03PRE2}. These so-called tilted-pulse fronts (TPFs) \cite{Fulop10Review} have a broad range of applications in nonlinear optics \cite{Martinez89IEEE,Szabo90APB,Dubietis97OL,DiTrapani98PRL,Liu00PRE,Wise02OPN,Schober07JOSAB}, quantum optics \cite{Torres05PRA,Hendrych09PRA}, and in the generation of terahertz radiation \cite{Hebling02OE,Hebling08JOSAB,Wang20LPR}. Nevertheless, angular dispersion offers only limited control over GVD; for example, in free space it always introduces anomalous GVD along the propagation direction of the pulse \cite{Martinez84JOSAA,Szatmari96OL}. This constraint also applies to previously reported spatially structured wave packets (e.g., pulsed Bessel beams \cite{Liu98JMO,Hu02JOSAA,Lu03JOSAA} and modified X-waves \cite{Sonajalg96OL,Sonajalg97OL}). These consequences can all be accounted for using a perturbative treatment of the propagation angle with respect to the optical frequency \cite{Martinez84JOSAA,Porras03PRE2}, which therefore presumes the \textit{differentiability} of the angular dispersion. Such a premise seems completely reasonable, especially for small angles and narrow bandwidths, and is usually taken for granted.

We recently demonstrated that a broad family of pulsed beams that we denote `space-time' (ST) wave packets \cite{Kondakci16OE,Parker16OE,Kondakci17NP,Yessenov19OPN}, in which the spatial frequencies and wavelengths are tightly associated \cite{Donnelly93ProcRSLA,Saari04PRE,Longhi04OE,Valtna07OC,Wong17ACSP1,Wong17ACSP2,Porras17OL,Efremidis17OL,Wong20AS}, is undergirded by a particular form of angular dispersion that is -- surprisingly -- \textit{non-differentiable} at the carrier frequency \cite{Hall21arxiv}. As a result, some consequences associated with conventional differentiable angular dispersion can be sidestepped by ST wave packets. First, the `universal' relationship that relates the pulse-front tilt to angular dispersion \cite{Hebling96OQE,Fulop10Review}, which is device-independent and also independent of the pulse bandwidth and shape, is violated by ST wave packets whose pulse-front tilt is proportional to the square root of the bandwidth \cite{Hall21arxiv}. Second, despite the underlying angular dispersion, ST wave packets are propagation-invariant \cite{Kondakci17NP,Bhaduri18OE,Bhaduri19OL,Schepler20ACSP,Yessenov20NC,Shiri20NC}, and can travel GVD-free at arbitrary group velocities \cite{Salo01JOA,Kondakci19NC,Bhaduri19Optica,Yessenov19OE,Bhaduri20NP}.

Here, we demonstrate that the unique non-differentiable angular dispersion intrinsic to ST wave packets facilitates exercising unprecedented control over the GVD they encounter in free space. By sculpting the spatio-temporal spectrum, \textit{independent control can be exercised over the magnitude and sign of all orders of dispersion}, which encompass the group velocity, second-order GVD, and higher-order dispersion terms -- thereby producing in principle an arbitrary dispersion profile in free space. Indeed, any dispersion order of the ST wave packet can be isolated and addressed separately from the others, so as to suppress or accentuate its coefficient across the pulse bandwidth while retaining diffraction-free axial evolution of the time-averaged intensity (or energy) \cite{Porras03PRE2}. This is in stark contrast to conventional optical materials, photonic structures, and previously studied structured optical fields in which the different dispersion orders are \textit{not} physically independent of each other. Uniquely, such control over the dispersion profile can be realized in any spectral band using the same experimental approach \cite{Kondakci18OE,Yessenov20OSAC}. Therefore, engineering the dispersion of freely propagating ST wave packets effectively renders free space an artificial medium with arbitrary dispersive characteristics. 

\section{Theory of angular dispersion and tilted pulse fronts}

\begin{figure*}[t!]
\centering
\includegraphics[width=16cm]{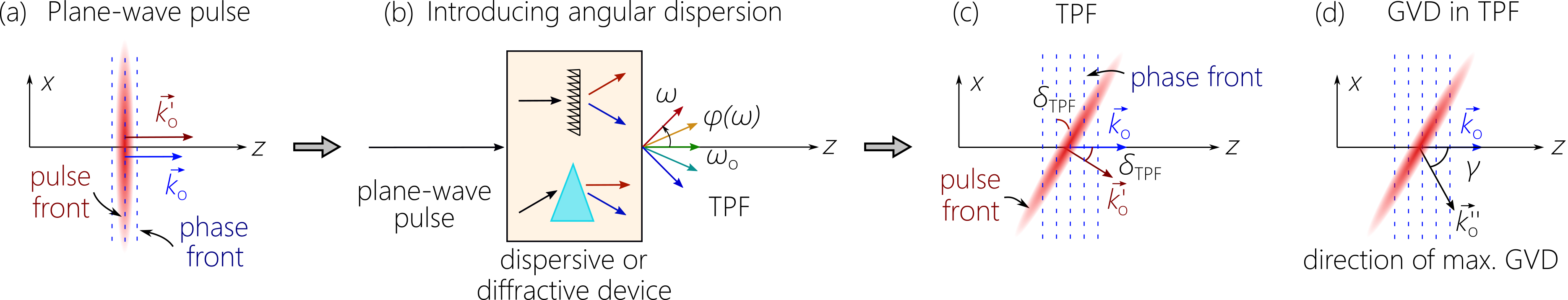}
\caption{Introducing differentiable angular dispersion into a plane-wave pulse. (a) The phase and pulse fronts of a plane-wave pulse are parallel, and both are orthogonal to the vectors $\vec{k}_{\mathrm{o}}$ and $\vec{k}_{\mathrm{o}}'$, which are parallel and coincide with the $z$-axis. (b) Angular dispersion is introduced into the field after the plane-wave pulse traverses a diffractive or dispersive device. The $z$-axis coincides with the direction of propagation of the frequency $\omega\!=\!\omega_{\mathrm{o}}$, $\varphi(\omega_{\mathrm{o}})\!=\!0$. (c) In the resulting TPF, the phase front is orthogonal to $\vec{k}_{\mathrm{o}}$ that coincides with the $z$-axis, whereas the pulse front is orthogonal to $\vec{k}_{\mathrm{o}}'$, which is tilted by an angle $\delta_{\mathrm{TPF}}$ with respect to $\vec{k}_{\mathrm{o}}$. (d) The direction of maximum GVD is along the vector $\vec{k}_{\mathrm{o}}''$, which is tilted by an angle $\gamma$ with respect to $\vec{k}_{\mathrm{o}}$.}
\label{Fig:CoventionalAngularDispersion}
\end{figure*}

We first describe briefly the conventional theory of angular dispersion \cite{Porras03PRE2,Torres10AOP,Fulop10Review}, starting with a scalar plane-wave pulse whose electrical field is given by $E(\vec{r};t)\!=\!\int\!d\omega\widetilde{E}(\omega)e^{i(kz-\omega t)}$, where $\widetilde{E}(\omega)$ is the Fourier transform of $E(0;t)$, $\omega$ is the temporal frequency, $k\!=\!\omega/c$ is the wave number, $c$ is the speed of light in vacuum, $\vec{r}$ is the position vector, and $z$ is the axial coordinate [Fig.~\ref{Fig:CoventionalAngularDispersion}(a)]. Angular dispersion is induced after traversing a diffractive or dispersive device (e.g., a grating or a prism), whereupon each frequency $\omega$ travels at a different angle $\varphi(\omega)$ [Fig.~\ref{Fig:CoventionalAngularDispersion}(b)] and the electric field takes the form:
\begin{equation}\label{Eq:BasicFieldEquation}
E(\vec{r};t)=\int\!d\omega\,\widetilde{E}(\omega)\,e^{i\xi(\vec{r};\omega)}e^{-i\omega t},
\end{equation}
where $\xi(\vec{r};\omega)\!=\!\vec{k}(\omega)\cdot\vec{r}$, and $\vec{k}(\omega)$ is the wave vector in free space, so that $|\vec{k}|\!=\!\omega/c$. A single integration over $\omega$ in Eq.~\ref{Eq:BasicFieldEquation} suffices because specifying the frequency $\omega$ identifies the wave vector $\vec{k}$ when the field is cylindrically symmetric or when it is held uniform along one transverse dimension (say $y$). Without loss of generality, we make use of the latter assumption for simplicity, so that $\vec{r}$ is restricted to the axial coordinate $z$ and the transverse coordinate $x$, and $\vec{k}\!=\!k(\sin{\varphi}\,\hat{x}+\cos{\varphi}\,\hat{z})$.

We denote the optical carrier frequency $\omega_{\mathrm{o}}$ and take its direction to coincide with the $z$-axis [Fig.~\ref{Fig:CoventionalAngularDispersion}(b)], and then expand the propagation angle to second order in $\Omega\!=\!\omega-\omega_{\mathrm{o}}$,
\begin{equation}\label{Eq:ExpansionForPhi}
\varphi(\omega)=\varphi(\omega_{\mathrm{o}}+\Omega)\approx\varphi_{\mathrm{o}}+\varphi_{\mathrm{o}}'\Omega+\tfrac{1}{2}\varphi_{\mathrm{o}}''\Omega^{2},
\end{equation}
where $\varphi_{\mathrm{o}}\!=\!\varphi(\omega_{\mathrm{o}})\!=\!0$, $\varphi_{\mathrm{o}}'\!=\!\tfrac{d\varphi}{d\omega}|_{\omega=\omega_{\mathrm{o}}}$, and $\varphi_{\mathrm{o}}''\!=\!\tfrac{d^{2}\varphi}{d\omega^{2}}|_{\omega=\omega_{\mathrm{o}}}$. The subscript `o' indicates quantities evaluated at $\omega\!=\!\omega_{\mathrm{o}}$, and primes denote derivatives with respect to $\omega$ \cite{Porras03PRE2}. Making use of Eq.~\ref{Eq:ExpansionForPhi}, noting that $k(\omega)\!=\!k_{\mathrm{o}}+\Omega/c$ in free space, and expanding $\sin{\varphi}$ and $\cos{\varphi}$ to second order in $\Omega$, we can expand the spectral phase $\xi(\vec{r};\omega)$ in Eq.~\ref{Eq:BasicFieldEquation} as:
\begin{equation}
\xi(\vec{r};\omega)\approx\xi_{\mathrm{o}}(\vec{r})+\xi_{\mathrm{o}}'(\vec{r})\Omega+\tfrac{1}{2}\xi_{\mathrm{o}}''(\vec{r})\Omega^{2}.
\end{equation}

The first term in this expansion is $\xi_{\mathrm{o}}(\vec{r})=\xi(\vec{r},\omega_{\mathrm{o}})=\vec{k}_{\mathrm{o}}\cdot\vec{r}\!=\!k_{\mathrm{o}}z$, so the \textit{phase front} (plane of constant phase) is orthogonal to the $z$-axis ($\vec{k}_{\mathrm{o}}\!=\!k_{\mathrm{o}}\hat{z}$), and the phase velocity is $c$. The second term is $\xi_{\mathrm{o}}'(\vec{r})=\vec{k}_{\mathrm{o}}'\cdot\vec{r}=k_{\mathrm{o}}\varphi_{\mathrm{o}}'x+z/c$, where the \textit{pulse front} (planes of constant amplitude) is orthogonal to $\vec{k}_{\mathrm{o}}'$. Such a field structure is known as a tilted pulse front (TPF) \cite{Fulop10Review}, and $\vec{k}_{\mathrm{o}}'$ is tilted by an angle $\delta_{\mathrm{TPF}}$ with respect to $\vec{k}_{\mathrm{o}}$ \cite{Martinez84JOSAA,Porras03PRE2} as shown in Fig.~\ref{Fig:CoventionalAngularDispersion}(c):
\begin{equation}\label{Eq:TPFTiltAngle}
\tan{\delta}_{\mathrm{TPF}}=\omega_{\mathrm{o}}\varphi_{\mathrm{o}}'.
\end{equation}
This universal relationship for the pulse-front tilt is a defining characteristic of TPFs \cite{Fulop10Review}. The third term in the expansion of $\xi(\vec{r};\omega)$ is:
\begin{equation}\label{Eq:GVDPhase}
\xi_{\mathrm{o}}''(\vec{r})=\vec{k}_{\mathrm{o}}''\cdot\vec{r}=(k_{\mathrm{o}}\varphi_{\mathrm{o}}''+\tfrac{2}{c}\varphi_{\mathrm{o}}')x-k_{\mathrm{o}}\varphi_{\mathrm{o}}'^{2}z,
\end{equation}
where $\vec{k}_{\mathrm{o}}''$ makes an angle $\gamma$ with respect to $\vec{k}_{\mathrm{o}}$ [Fig.~\ref{Fig:CoventionalAngularDispersion}(d)], and
\begin{equation}
\tan{\gamma}=-\left\{\frac{\varphi_{\mathrm{o}}''}{\varphi_{\mathrm{o}}'^{2}}+2\cot{\delta}\right\}.
\end{equation}
Maximal GVD occurs along $\vec{k}_{\mathrm{o}}''$, and the field has constant spectral chirp in planes orthogonal to it \cite{Porras03PRE2}. We are typically interested in the GVD coefficient along the $z$-axis (the direction of propagation), given here by: 
\begin{equation}\label{Eq:TPFAnomalousGVD}
k_{\mathrm{o},z}''=-k_{\mathrm{o}}\varphi_{\mathrm{o}}'^{2},
\end{equation}
which is always negative-valued in free space, thus corresponding to \textit{anomalous} GVD \cite{Martinez84JOSAA,Szatmari96OL,Porras03PRE2}.

\begin{figure*}[t!]
\centering
\includegraphics[width=14cm]{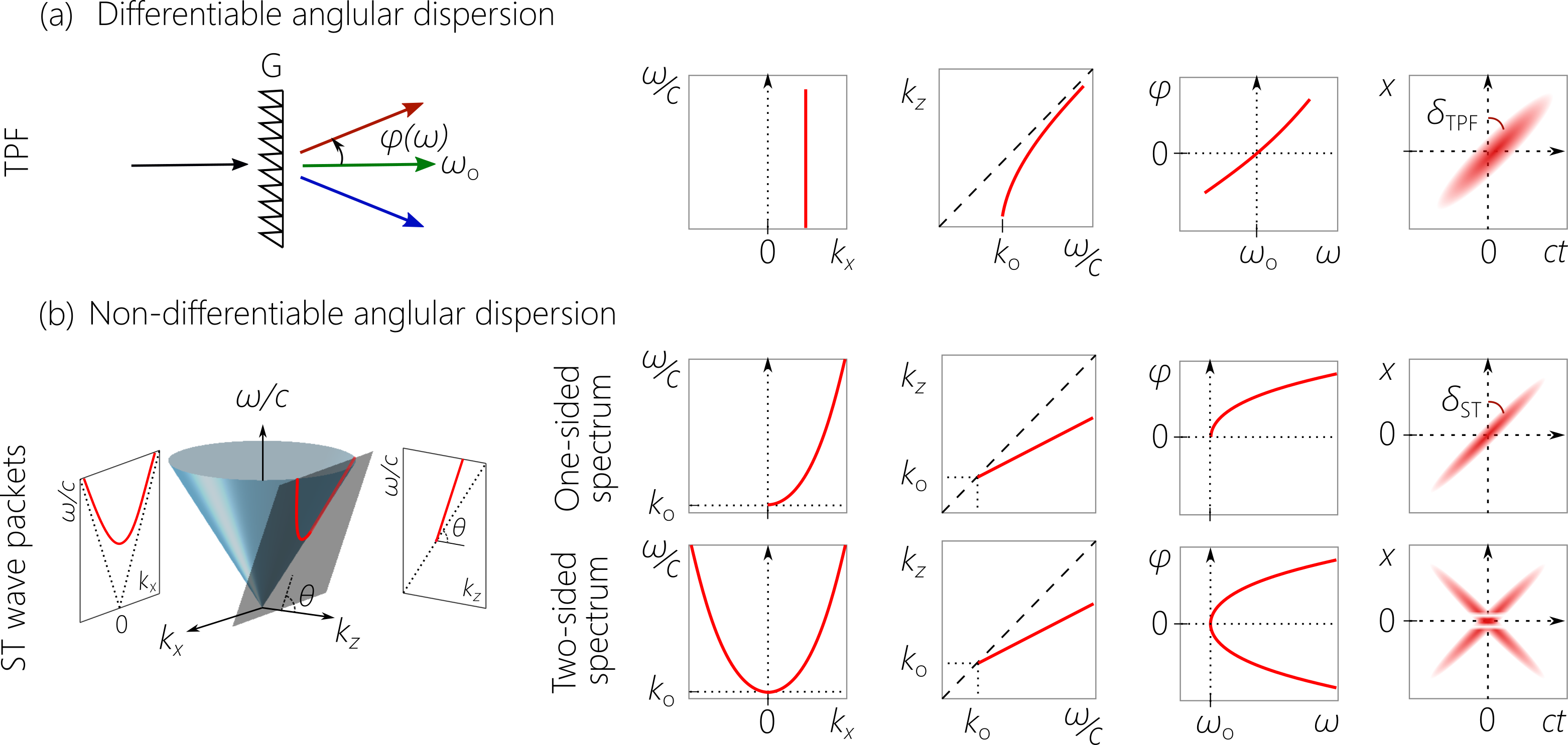}
\caption{(a) Differentiable angular dispersion underlying a TPF. Angular dispersion is introduced into a plane-wave pulse incident normally on a grating, leading to a tilt in the pulse front. Shown are the spectral projections onto the $(k_{x},\tfrac{\omega}{c})$ and $(k_{z},\tfrac{\omega}{c})$ planes, the angular dispersion $\varphi(\omega)$, and the spatio-temporal profile $I(x,z;t)\!=\!|\psi(x,z;t)|^{2}$ at $z\!=\!0$. (b) Non-differentiable angular dispersion underlying ST wave packets. By constraining the spectral support domain on the surface of the light-cone to a 1D trajectory whose projection onto the $(k_{z},\tfrac{\omega}{c})$-plane is a straight line. The first row shows a ST wave packet with a one-sided spatial spectrum (positive-valued $k_{x}$), while the second row shows the two-sided spatial spectrum counterpart having a symmetrized spatio-temporal profile.}
\label{Fig:STWavePacketsAndAngularDispersion}
\end{figure*}

We can now write the field as a carrier and a slowly varying envelope, $E(x,z;t)\!=\!e^{i(k_{\mathrm{o}}z-\omega_{\mathrm{o}}t)}\psi(x,z;t)$, where
\begin{equation}
\psi(x,z;t)=\int\!d\Omega\,\widetilde{\psi}(\Omega)e^{-i\{t-\xi_{\mathrm{o}}'(\vec{r})\}\Omega}e^{i\xi_{\mathrm{o}}''(\vec{r})\Omega^{2}/2},
\end{equation}
and $\widetilde{\psi}(\Omega)$ is the Fourier transform of $\psi(0,0;t)$. It is clear that the group velocity is determined by the term $\xi_{\mathrm{o}}'(\vec{r})$, and the propagation is dispersive in free space because of the term $\xi_{\mathrm{o}}''(\vec{r})$.

In summary, introducing angular dispersion into a pulsed field leads to a tilt in the pulse front (Eq.~\ref{Eq:TPFTiltAngle}) and anomalous GVD along the propagation direction (Eq.~\ref{Eq:TPFAnomalousGVD}). Expanding $\varphi(\omega)$ in a Taylor series around $\omega\!=\!\omega_{\mathrm{o}}$ (Eq.~\ref{Eq:ExpansionForPhi}) is the starting point for this analysis. This appears to be a valid premise, especially for small angles and narrow bandwidths. A standard example is provided in Fig.~\ref{Fig:STWavePacketsAndAngularDispersion}(a), where a plane-wave pulse after normal incidence on a grating has a constant transverse wave number $k_{x}$ and $\varphi(\omega_{\mathrm{o}}+\Omega)\!\propto\!\Omega$, thus resulting in a TPF structure and anomalous GVD. We proceed to show that these findings are all fundamentally overturned when considering instead ST wave packets that are undergirded by non-differentiable angular dispersion.

\section{Theory of angular dispersion for space-time wave packets}

\subsection{Dispersion-free ST wave packets}

ST wave packets are pulsed beams endowed with a precise spatio-temporal structure that renders them propagation-invariant (i.e., diffraction-free and dispersion-free) \cite{Reivelt03arxiv,Kiselev07OS,Turunen10PO,FigueroaBook14}. Underlying these ST wave packets is a particular form of angular dispersion, which ensures that the axial wave number $k_{z}$ satisfies the constraint $k_{z}(\Omega)\!=\!k_{\mathrm{o}}+\Omega/\widetilde{v}$, where $\widetilde{v}$ is the group velocity \cite{Turunen10PO,Kondakci17NP,Yessenov19PRA}. This is the equation of a plane $\mathcal{P}(\theta)$ that is parallel to the $k_{x}$-axis and tilted by an angle $\theta$ with respect to the $k_{z}$-axis, such that $\widetilde{v}\!=\!c\tan{\theta}$. Consequently, the support domain of a ST wave packet in the spectral space $(k_{x},k_{z},\tfrac{\omega}{c})$ is a conic section at the intersection of the free-space light-cone $k_{x}^{2}+k_{z}^{2}\!=\!(\tfrac{\omega}{c})^{2}$ with the plane $\mathcal{P}(\theta)$ \cite{Yessenov19PRA,Yessenov19OPN}. As is clear in Fig.~\ref{Fig:STWavePacketsAndAngularDispersion}(b), each spatial frequency $k_{x}$ is associated with a single temporal frequency $\omega$, and their relationship for narrow bandwidths takes approximately the form of a parabola \cite{Porras17OL,Kondakci17NP,Bhaduri20NP}. If only positive values of $k_{x}$ are considered (one-sided spatial spectrum), then the wave packet has the form of a TPF with a tilt angle $\delta_{\mathrm{ST}}$ \cite{Kondakci19ACSP,Hall21arxiv}. When both positive and negative values of $k_{x}$ are considered (two-sided spatial spectrum), the wave packet has a symmetrized TPF structure.

Starting with $k_{x}^{2}\!=\!k^{2}-k_{z}^{2}$ and substituting $k_{z}\!=\!k_{\mathrm{o}}+\Omega/\widetilde{v}$, $k\!=\!k_{\mathrm{o}}+\Omega/c$, and $k_{x}\!=\!k\sin{\varphi}$, we obtain:
\begin{equation}\label{Eq:SinPhiForGVDFree}
\sin{\{\varphi(\Omega)\}}\approx\eta\frac{\sqrt{\Omega/\omega_{\mathrm{o}}}}{1+\Omega/\omega_{\mathrm{o}}},
\end{equation}
where $\eta^{2}\!=\!2|1-\cot{\theta}|$. In the narrowband regime $\Omega/\omega_{\mathrm{o}}\!\ll\!1$, we have $\sin{\varphi}\!\propto\!\sqrt{\Omega}$, which is \textit{not} differentiable at $\Omega\!=\!0$, even for small angles, so that a perturbative expansion (Eq.~\ref{Eq:ExpansionForPhi}) is \textit{not} justified. Therefore, the conventional theory outlined above is not applicable because the derivatives $\varphi_{\mathrm{o}}'$ and $\varphi_{\mathrm{o}}''$ are not well-defined. This non-differentiability leads to several unique features that depart from those of conventional TPFs. First, the pulse-front tilt angle $\delta_{\mathrm{ST}}$ does \textit{not} follow the universal relationship in Eq.~\ref{Eq:TPFTiltAngle}, and is instead determined by a new formula, $\tan{\delta}_{\mathrm{ST}}\!=\!\pm\tfrac{\eta}{2}\tfrac{1}{\sqrt{\Delta\Omega/\omega_{\mathrm{o}}}}$ \cite{Hall21arxiv}. Surprisingly, $\delta_{\mathrm{ST}}$ depends on the pulse bandwidth $\Delta\Omega$, in addition to the frequency-independent spectral tilt angle $\theta$. Second, ST wave packets are GVD-free in free space despite the strong angular dispersion (Eq.~\ref{Eq:GVDPhase} does not hold) \cite{Yessenov19OE,Wong20AS,Wong17ACSP2}. Indeed, the envelope of the ST wave packet takes the form:
\begin{equation}
\psi(x,z;t)\!=\!\!\int\!d\Omega\,\widetilde{\psi}(\Omega)e^{ik_{x}(\Omega)x}e^{-i(t-z/\widetilde{v})\Omega}\!=\!\psi(x,0,t\!-\!z/\widetilde{v}),
\end{equation}
which corresponds to rigid transport at a group velocity $\widetilde{v}\!=\!c\tan{\theta}$. Third, along the propagation axis $z$, the group velocity $\widetilde{v}$ can be tuned over an unprecedented span while remaining in the paraxial regime (at least for narrow bandwidths) by changing the angular dispersion to adjust $\theta$ \cite{Kondakci19NC}.

\subsection{Arbitrary GVD in free space using ST wave packets}

\begin{figure*}[t!]
\centering
\includegraphics[width=16cm]{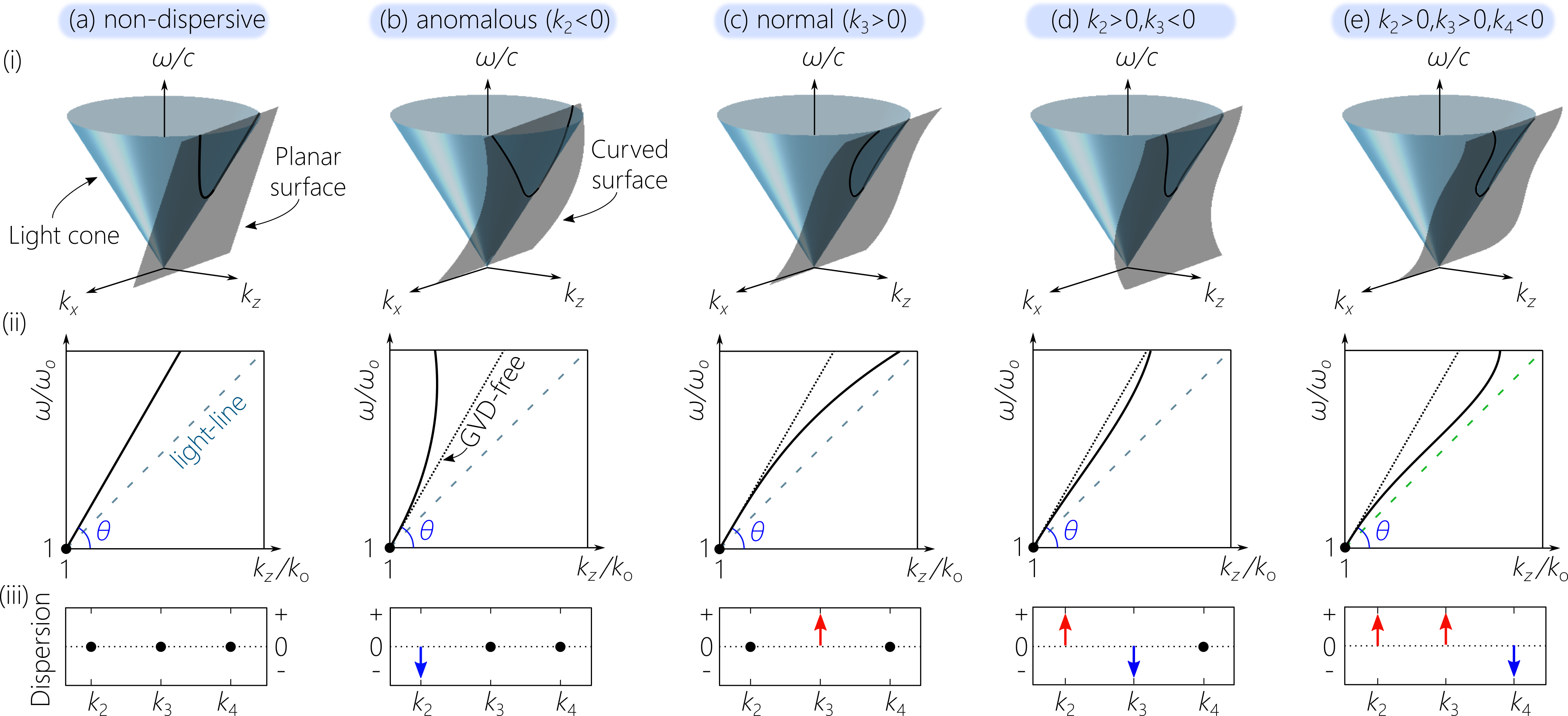}
\caption{Concept of GVD control for ST wave packets in free space. (a) Non-dispersive ST wave packets; (b) dispersive ST wave packets with negative $k_{2}$ (anomalous); (c) positive $k_{3}$ (normal); (d) combination of positive $k_{2}$ and negative $k_{3}$; (e) combination of all three orders of GVD: positive $k_{2}$, positive $k_{3}$, and negative $k_{4}$. Row (i) depicts the light-cone intersecting with a spectral hypersurface perpendicular to $(k_{z},\tfrac{\omega}{c})$-plane; row (ii) shows the spectral projection onto the $(k_{z},\tfrac{\omega}{c})$-plane; and row (iii) depicts the selected dispersion orders and the sign of each for each column. Dashed lines are the light-line, dotted lines correspond to the ST wave packet are eliminating dispersion, and the solid curves to the dispersive ST wave packets.}
\label{Fig:Concept}
\end{figure*}

The non-differentiable angular dispersion underpinning ST wave packets allows us to circumvent the fundamental constraint in Eq.~\ref{Eq:TPFAnomalousGVD}. Specifically, \textit{arbitrary} dispersion profiles can be realized, including prescribed magnitude \textit{and} sign for the dispersion coefficients of all orders, by synthesizing ST wave packets whose spectral support domain on the light-cone lies at its intersection \textit{not} with a plane [Fig.~\ref{Fig:Concept}(a)], but instead with a surface of the form [Fig.~\ref{Fig:Concept}(b-e)]:
\begin{equation}\label{Eq:CurvedProjection}
k_{z}=k_{\mathrm{o}}+f(\Omega).
\end{equation}
This is a planar curved surface that is parallel to the $k_{x}$-axis, and $f(\Omega)$ is assumed to be a polynomial function of $\Omega$,
\begin{equation}
f(\Omega)=k_{1}\Omega+\frac{1}{2}k_{2}\Omega^{2}+\cdots=\sum_{n=1}\frac{k_{n}}{n!}\Omega^{n},
\end{equation}
where $k_{1}\!=\!\tfrac{1}{\widetilde{v}}$. The projection of this surface onto the $(k_{z},\tfrac{\omega}{c})$-plane is thus the 1D curve in Eq.~\ref{Eq:CurvedProjection}. In principle, any desired dispersion profile can be realized by precisely tuning the values of the coefficients of the polynomial terms in $f(\Omega)$, which correspond physically to the weights of the various dispersion orders experienced by the ST wave packet in free space.

Examples are shown in Fig.~\ref{Fig:Concept} starting with $f(\Omega)\!=\!\Omega/\widetilde{v}$ for a propagation-invariant ST wave packet [Fig.~\ref{Fig:Concept}(a)]. Arbitrary dispersion profiles are realized by adding more terms to $f(\Omega)$: anomalous GVD with $f(\Omega)\!=\!\Omega/\widetilde{v}-\tfrac{1}{2}|k_{2}|\Omega^{2}$ [Fig.~\ref{Fig:Concept}(b)]; normal third-order GVD with $f(\Omega)\!=\!\Omega/\widetilde{v}+\tfrac{1}{6}|k_{3}|\Omega^{3}$ [Fig.~\ref{Fig:Concept}(c)]; a combination of normal second-order and anomalous third-order dispersion terms with $f(\Omega)\!=\!\Omega/\widetilde{v}+\tfrac{1}{2}|k_{2}|\Omega^{2}-\tfrac{1}{6}|k_{3}|\Omega^{3}$ [Fig.~\ref{Fig:Concept}(d)]; and a combination of second-, third-, and fourth-order dispersion terms [Fig.~\ref{Fig:Concept}(e)]. In any of these cases, the desired form of $f(\Omega)$ dictates the necessary angular dispersion to be introduced into the field according to:
\begin{eqnarray}\label{Eq:GeneralAngleForDispersiveSTWP}
\sin^{2}{\{\varphi(\Omega)\}}&=&1-\left(\frac{1+f(\Omega)/k_{\mathrm{o}}}{1+\Omega/\omega_{\mathrm{o}}}\right)^{2}\nonumber\\&\approx&2\frac{\Omega/\omega_{\mathrm{o}}}{(1+\Omega/\omega_{\mathrm{o}})^{2}}\left(1-c\sum_{n=1}\frac{k_{n}}{n!}\Omega^{n-1}\right).
\end{eqnarray}
This formula is the basis for synthesizing dispersive ST wave packets using the two-step spectral synthesis procedure outlined below. To the best of our knowledge, dispersion profiles such as those shown in Fig.~\ref{Fig:Concept}(c-e) have not been demonstrated heretofore.

\subsection{Special case: Normal and anomalous second-order GVD}

\begin{figure*}[t!]
\centering
\includegraphics[width=16cm]{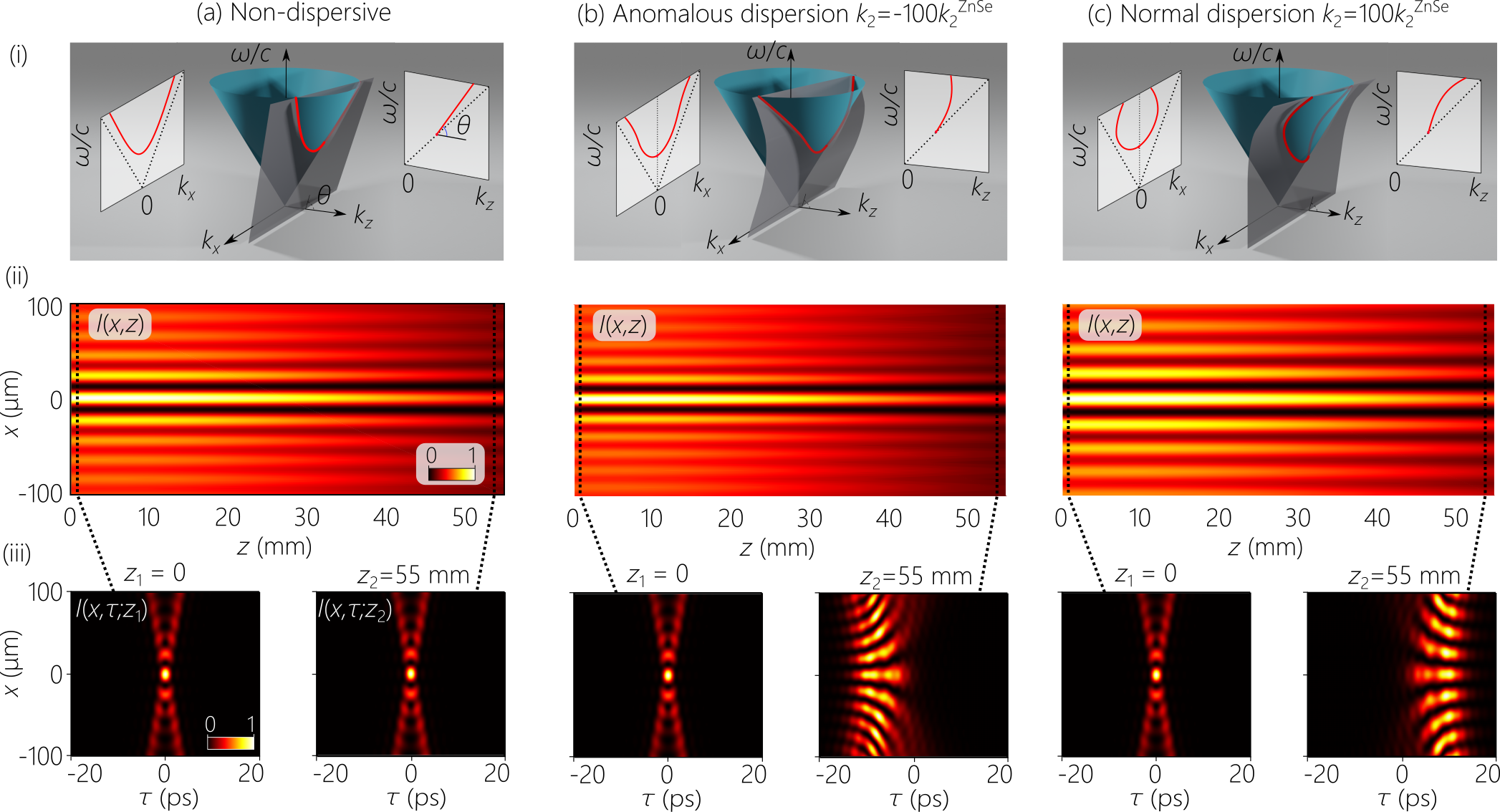}
\caption{Calculated characteristics of nondispersive and dispersive ST wave packets with second-order GVD. (a) Nondispersive ST wave packets with a spectral tilt angle $\theta=50^{\circ}$; (b) ST wave packets with anomalous dispersion ($k_{2}=-100k_{2}^{\mathrm{ZnSe}}$) and (c) normal dispersion ($k_{2}=100k_{2}^{\mathrm{ZnSe}}$), where $k_{2}^{\mathrm{ZnSe}}$ is GVD of ZnSe. (i) Spatio-temporal spectrum of ST wave packets at the intersection of the light-cone with a spectral hypersurface. (ii) Time-averaged intensity profile $I(x,z)$. (iii) Time-resolved intensity profiles at two axial locations $z=0$ and $z=55$~mm. The bandwidth throughout is $\Delta\lambda\!\approx\!1$~nm.}
\label{Fig:SecondOrderGVD}
\end{figure*}

Consider the simplest case of GVD in which $k_{n}\!=\!0$ for $n\!\geq\!3$, so that $k_{z}\!=\!k_{\mathrm{o}}+\Omega/\widetilde{v}+\tfrac{1}{2}k_{2}\Omega^{2}$, then the on-axis ($x\!=\!0$) envelope is
\begin{equation}
\psi(x,z,t)=\int\!d\Omega\widetilde{\psi}(\Omega)e^{-i(t-z/\widetilde{v})\Omega}e^{ik_{2}\Omega^{2}z/2},
\end{equation}
which has the form of a plane-wave pulse traveling in a medium with a group index $c/\widetilde{v}$ and GVD parameter $k_{2}$. Unlike the anomalous GVD induced by conventional differentiable angular dispersion \cite{Martinez84JOSAA,Porras03PRE2}, $k_{2}$ may take on positive or negative values, corresponding to either normal or anomalous GVD, respectively. We plot the spectral support domain on the light-cone and the spectral projections onto the $(k_{x},\tfrac{\omega}{c})$ and $(k_{z},\tfrac{\omega}{c})$ planes for a propagation-invariant ST wave packet ($k_{2}\!=\!0$) in Fig.~\ref{Fig:SecondOrderGVD}(a), for a dispersive ST wave packet undergoing anomalous GVD in Fig.~\ref{Fig:SecondOrderGVD}(b), and undergoing normal GVD in Fig.~\ref{Fig:SecondOrderGVD}(c). In all cases, the time-averaged intensity $I(x,z)\!=\!\int\!dt|E(x,z,t)|^{2}$ (or energy) is nevertheless diffraction-free:
\begin{equation}
I(x,z)\!=\!\!\int\!\!\!\int\!\!dk_{x}dk_{x}'\,\widetilde{\psi}(k_{x})\widetilde{\psi}^{*}(k_{x})e^{i(k_{x}-k_{x}')x}\delta(|k_{x}|\!-\!|k_{x}'|)=I(x,0),
\end{equation}
which is independent of $z$ altogether \cite{Porras03PRE2,Yessenov19Optica,Yessenov19OL}. Time-resolved measurements should, however, reveal that the underlying spatio-temporal intensity profile of the wave packets disperse when $k_{2}\!\neq\!0$. 

\section{Experimental setup}

\begin{figure}[t!]
\centering
\includegraphics[width=8.6cm]{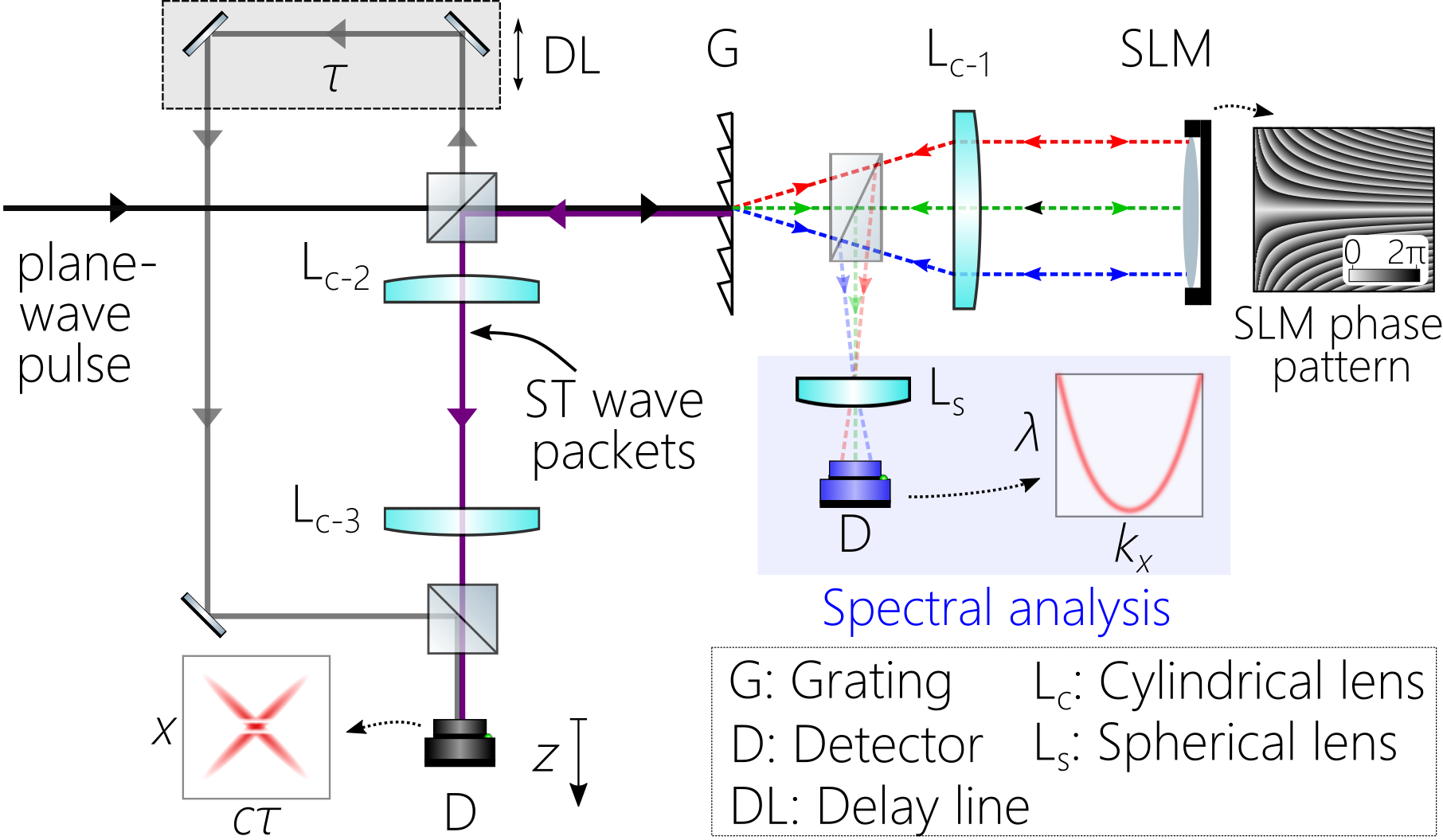}
\caption{Setup for synthesis and characterization of ST wave packets with controllable dispersion. Inset (right) shows the phase distribution implemented on the SLM to realize a ST wave packet. We provide a key for the optical components below the schematic.}
\label{Fig:Setup}
\end{figure}

Whereas differentiable angular dispersion can be inculcated by a conventional diffractive or dispersive device, its non-differentiable counterpart is introduced via a two-step procedure capable of inducing \textit{arbitrary} angular dispersion into a plane-wave pulse as shown in Fig.~\ref{Fig:Setup}  \cite{Yessenov19OPN,Yessenov19PRA,Yessenov19OE}. In the first step, a grating and a collimating lens spatially resolve the spectrum, but we do \textit{not} rely on the grating to inculcate the angular dispersion. Instead, we make use of a spatial light modulator (SLM) that imprints upon the impinging spectrally resolved wave front a two-dimensional phase distribution designed to associate a prescribed propagation angle $\varphi(\omega)$ with each frequency $\omega$. In this way, an arbitrary non-differentiable functional form for $\varphi(\omega)$ can be produced. The phase-modulated wave front is then retro-reflected from the SLM through the lens back to the grating that reconstitutes the pulse and produces the ST wave packet. Alternatively, one can make use of an unfolded layout with the SLM replaced by a phase plate, which is useful in applications that make use of high-energy laser pulses \cite{Kondakci18OE}, broadband spectra \cite{Bhaduri19OL}, or wavelengths for which SLMs are not available (e.g., in the mid-infrared \cite{Yessenov20OSAC}).

We carry out our experiments in the folded configuration depicted in Fig.~\ref{Fig:Setup} starting with $\sim\!100$~fs pulses at a central wavelength of $\sim\!800$~nm from a mode-locked Ti:sapphire laser (Spectra Physics, Tsunami), which are directed to a diffraction grating ($1200$~lines/mm). The first diffraction order is collimated by a cylindrical lens of focal length 50~cm before normal incidence on a phase-only reflective SLM (Hamamatsu X10468-02). In the process of spectral phase modulation, we spectrally filter the pulse from the initial bandwidth of $\approx\!10$~nm ($\sim\!100$-fs pulsewidth) to $\approx\!1$~nm ($\sim\!1.5$-ps pulsewidth).

We characterize the synthesized ST wave packets in two domains. First, we measure the spatio-temporal spectrum by a combination of a grating that resolves the temporal spectrum and a lens in a $2f$-configuration that resolves the spatial spectrum. This yields the spatio-temporal spectrum projected onto the $(k_{x},\lambda)$-plane, from which we extract (1) the spectral projection onto the $(k_{z},\lambda)$-plane to ascertain that the desired constraint $k_{z}\!=\!k_{\mathrm{o}}+f(\Omega)$ has been satisfied \cite{Kondakci17NP}, and (2) the angular dispersion $\varphi(\lambda)$. Second, we obtain the spatio-temporal intensity profile $I(x,z;\tau)$ at different axial planes $z$ by interfering the ST wave packet with the original short pulse from the Ti:sapphire laser (after an optical delay $\tau$) at a detector placed at $z$. By monitoring the visibility of the spatial resolved fringes while sweeping $\tau$, we reconstruct the envelope $I(x,z;\tau)$. By repeating these measurements at different $z$, we can assess the impact of the induced GVD. The spatio-temporal profile measurements are recorded in a moving time-frame traveling at the group velocity $\widetilde{v}\!=\!c\tan{\theta}$.

\section{Measurements}

For each ST wave packet, we obtain 5 quantities: (1) the measured spectral intensity projected onto the $(k_{x},\lambda)$-plane, denoted $|\widetilde{\psi}(k_{x},\lambda)|^{2}$; (2) the extracted spectral intensity projected onto the $(k_{z},\lambda)$-plane, denoted $|\widetilde{\psi}(k_{z},\lambda)|^{2}$; (3) the extracted angular dispersion $\varphi(\lambda)$; (4) the spatio-temporal intensity profile $I(x,z;\tau)\!=\!|\psi(x,z;\tau)|^{2}$ at $z\!=\!0$; and (5) $I(x,z;\tau)$ at $z\!=\!55$~mm. This distance of 55~mm is selected to show clearly the impact of the induced GVD in all the cases examined, and $\tau$ denotes time measured in a frame traveling at $\widetilde{v}\!=\!c\tan{\theta}$.

\begin{figure*}[t!]
\centering
\includegraphics[width=16cm]{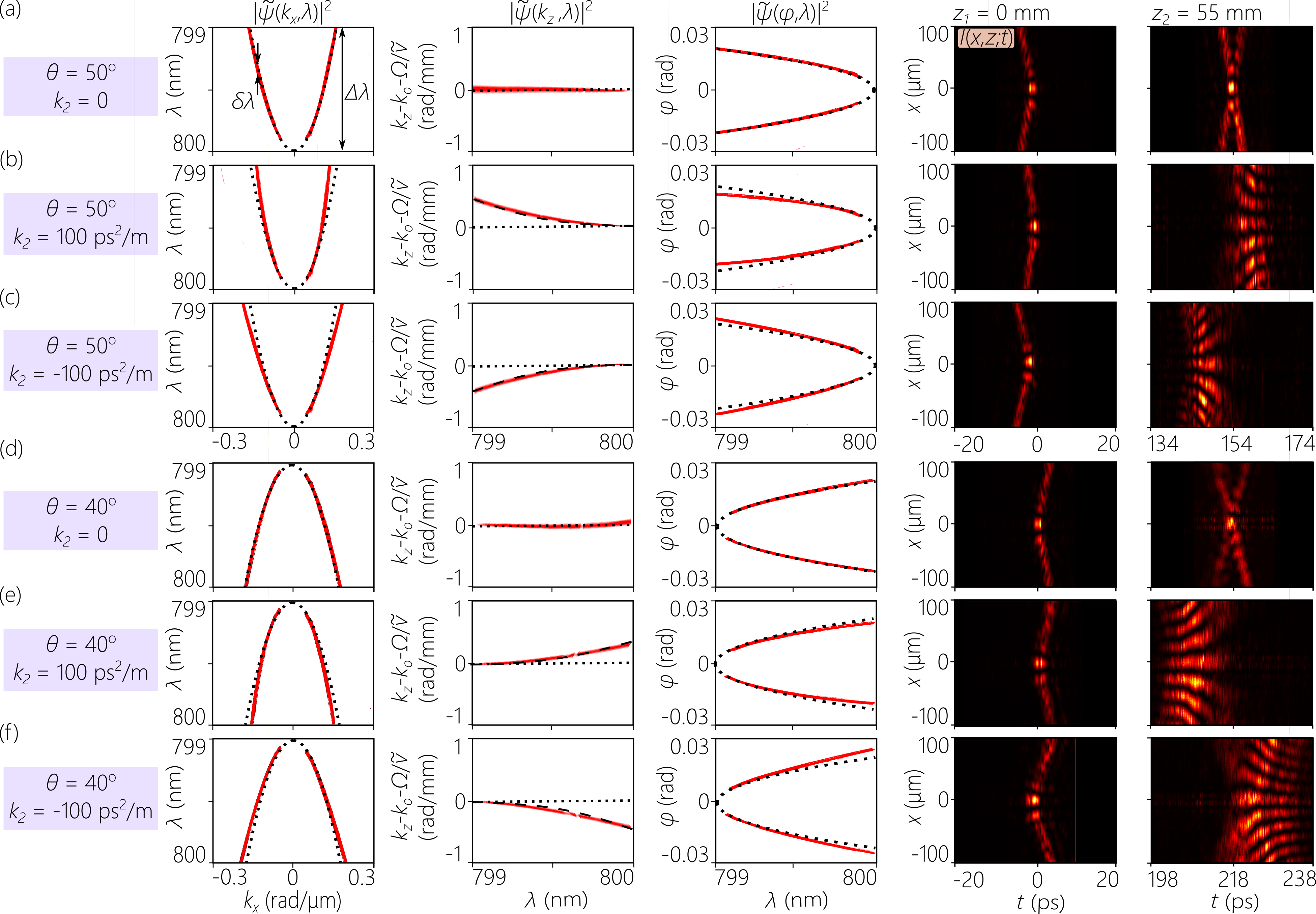}
\caption{Measurement results for dispersive ST wave packets with second-order GVD. In the first column we plot the spectral projection onto the $(k_{x},\lambda)$-plane; in the second the spectral projection onto the $(k_{z},\lambda)$-plane after shifting $k_{z}\!\rightarrow\!k_{z}-k_{\mathrm{o}}-\Omega/\widetilde{v}$ to isolate the higher-order dispersion terms; in the third the angular dispersion $\varphi(\lambda)$; and in the final two columns the spatio-temporal intensity profiles $I(x,z;t)$ at $z\!=\!0$ and at $z\!=\!55$~mm. Note that we plot the profiles with $t$ rather than in a moving frame $\tau$, so that the difference in group delay (related to the group velocity $\widetilde{v}\!=\!c\tan{\theta}$) is manifest. Throughout, the dotted curves are theoretical expectations for GVD-free ST wave packets, whereas the dashed curves in the second column represent the theoretical expectation for the dispersive wave packets. The panels in each row are arranged similarly to those in Fig.~\ref{Fig:STWavePacketsAndAngularDispersion}, except that the temporal frequency $\omega$ is replaced with the wavelength $\lambda$. (a-c) Superluminal ST wave packets with $\theta\!=\!50^{\circ}$ ($\widetilde{v}\!=\!1.19c$). (a) Propagation-invariant ST wave packet with $k_{2}\!=\!0$; (b) same as (a) for a dispersive ST wave packet with normal GVD, $k_{2}\!\approx\!100$~ps$^2$/m; and (c) same as (a) for a dispersive ST wave packet with anomalous GVD, $k_{2}\!=\!-100$~ps$^2$/m. In (a), we identify the full bandwidth $\Delta\lambda\!\approx\!1$~nm, and the spectral uncertainty $\delta\lambda\!\sim\!25$~pm. (d-f) Same as (a-c) for subluminal ST wave packets with $\theta\!=\!40^{\circ}$ ($\widetilde{v}\!=\!0.84c$).}
\label{Fig:Measurements_k2}
\end{figure*}

We first confirm the predicted normal and anomalous second-order GVD for ST wave packets in free space -- in contradistinction to the anomalous GVD associated solely with TPFs \cite{Martinez84JOSAA,Porras03PRE2,Torres10AOP}. We plot in Fig.~\ref{Fig:Measurements_k2}(a-c) measurements for three superluminal ST wave packets at $\theta\!=\!50^{\circ}$ ($\widetilde{v}\!\approx\!1.19c$). We first plot in Fig.~\ref{Fig:Measurements_k2}(a) the measurements for a propagation-invariant ST wave packet in which GVD is altogether suppressed $k_{2}\!=\!0$, so that $k_{z}\!=\!k_{\mathrm{o}}+\Omega/\widetilde{v}$. In this case, the angular dispersion conforms well to $\varphi(\Omega)\!\propto\!\sqrt{\Omega}$ \cite{Hall21arxiv}. We also plot the spectral projection onto the $(k_{z}\lambda)$-plane in terms of $k_{z}-k_{\mathrm{o}}-\Omega/\widetilde{v}\!=\!f(\Omega)-\Omega/\widetilde{v}$ to isolate terms higher than linear in $f(\Omega)$. Here, the spectral projection is flat in absence of GVD as expected, and the wave-packet profile remains invariant along the axial span from $z\!=\!0$ to 55~mm.

Next, we plot measurements for two ST wave packets having equal-magnitude GVD but of opposite signs, $k_{z}\!=\!k_{\mathrm{o}}+\Omega/\widetilde{v}+k_{2}\Omega^{2}/2$. In most transparent optical material away from its resonances, a $\sim\!1$-ps pulse will not experience appreciable GVD-induced pulse broadening over a distance of 55~mm. However, our strategy allows for introducing large values of GVD that may be readily set to significantly larger than those for conventional optical materials. We take as a reference material ZnSe whose GVD parameter is $k_{2}^{\mathrm{ZnSe}}\!\approx\!1054$~fs$^2$/mm at a wavelength of 800~nm. We select the GVD parameter for the ST wave packets here to be $k_{2}\!=\!\pm100k_{2}^{\mathrm{ZnSe}}$, and plot the results for normal GVD $k_{2}\!=\!100k_{2}^{\mathrm{ZnSe}}$ in Fig.~\ref{Fig:Measurements_k2}(b), and for anomalous GVD $k_{2}\!=\!-100k_{2}^{\mathrm{ZnSe}}$ in Fig.~\ref{Fig:Measurements_k2}(c). Note that the axial wave number $k_{z}-k_{\mathrm{o}}-\Omega/\widetilde{v}$ is no longer flat and equal to zero, and instead it curves upwards or downwards according to the sign of $k_{2}$. Similarly, the sign of $k_{2}$ determines whether the angular dispersion $\varphi(\lambda)$ is smaller or larger than that of the propagation-invariant ST wave packet. Although our measurement technique does not identify the sign of $k_{2}$ directly, the difference in the sign of GVD between Fig.~\ref{Fig:Measurements_k2}(b) and Fig.~\ref{Fig:Measurements_k2}(c) is nevertheless clear from the opposite direction of pulse broadening with respect to the propagation-invariant wave packet in Fig.~\ref{Fig:Measurements_k2}(a). 

\begin{figure*}[t!]
\centering
\includegraphics[width=16cm]{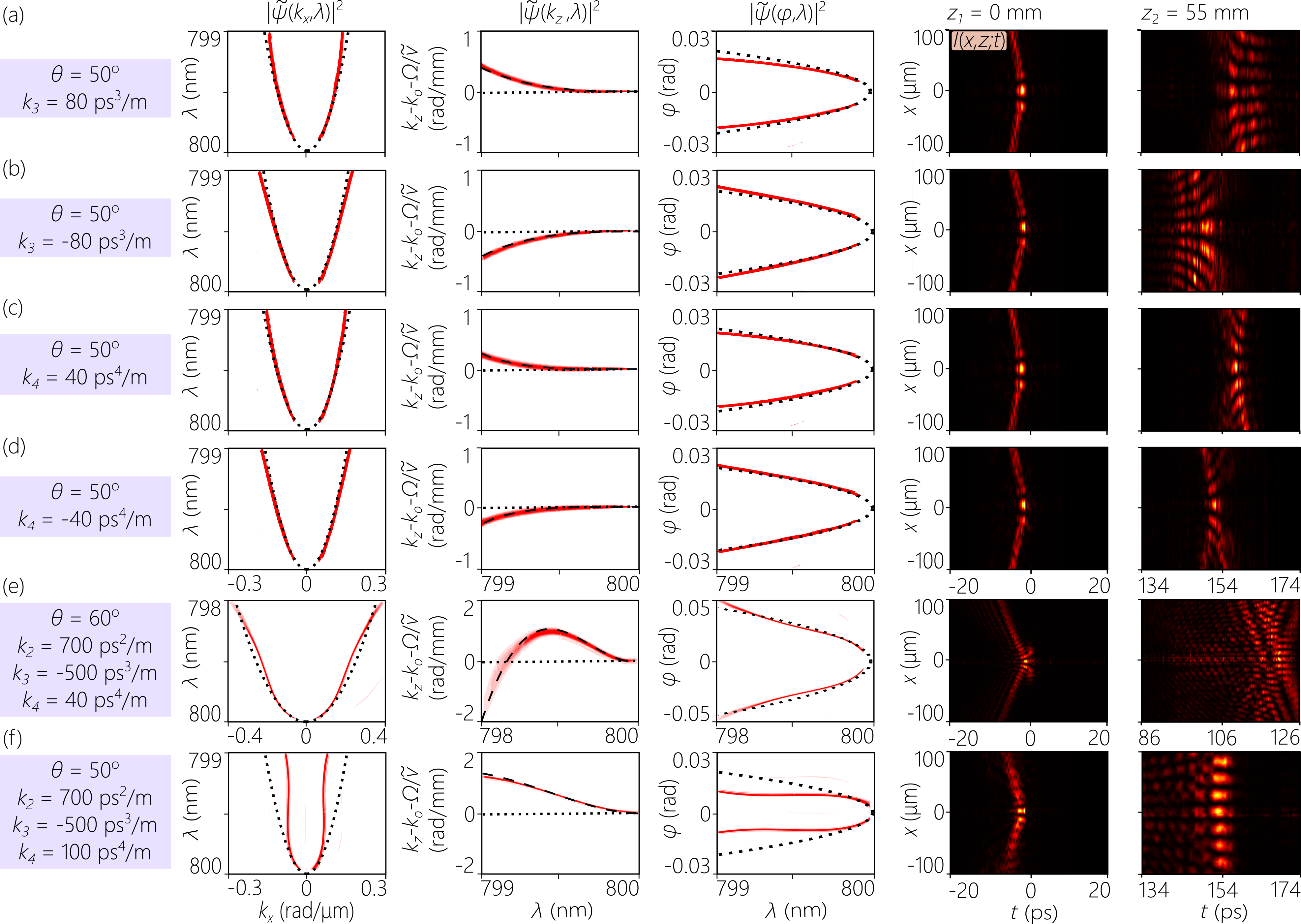}
\caption{Measurement results for dispersive ST wave packets with higher-order GVD terms. The arrangement of the panels and the significance of the dashed and dotted curves are all similar to Fig.~\ref{Fig:Measurements_k2}. (a,b) The spectral tilt angle is $\theta\!=\!50^{\circ}$ ($\widetilde{v}\!=\!1.19c$) and a third-order dispersion term is included with (a) $k_{3}\!=\!80$~ps$^{3}$/mm and (b) $k_{3}\!=\!-80$~ps$^{3}$/mm. (c,d) The spectral tilt angle is $\theta\!=\!50^{\circ}$ ($\widetilde{v}\!=\!1.19c$) and a fourth-order dispersion term is included with (c) $k_{4}\!=\!40$~ps$^{4}$/mm and (d) $k_{4}\!=\!-40$~ps$^{4}$/mm. (e,f) Dispersive ST wave packets with the first four dispersion terms ($\widetilde{v}$, $k_{2}$, $k_{3}$, and $k_{4}$) all having non-zero values.}
\label{Fig:Measurements_higherOrder}
\end{figure*}

\begin{figure*}[t!]
\centering
\includegraphics[width=16.0cm]{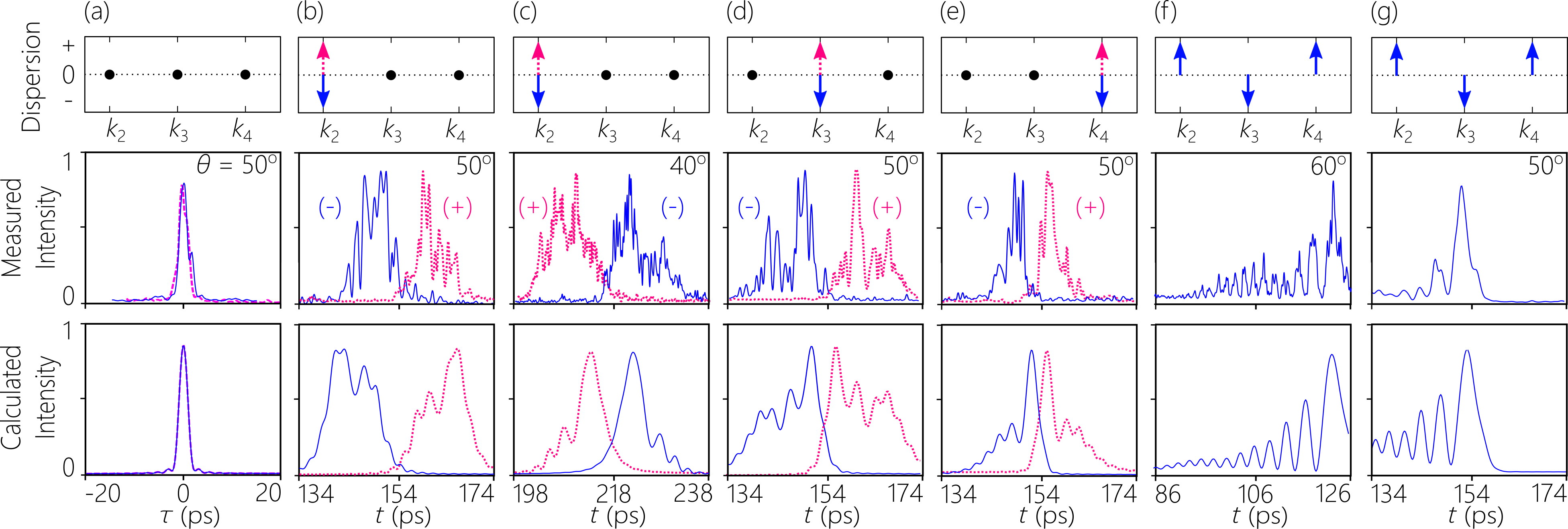}
\caption{Comparison of the on-axis temporal intensity profiles $I(x\!=\!0,z;t)$ from Fig.~\ref{Fig:Measurements_k2} and Fig.~\ref{Fig:Measurements_higherOrder} (middle row) with calculated profiles (bottom row) at $z\!=\!55$~mm. In each panel we indicate the spectral tilt angle $\theta$ that determines the group velocity. (a) The GVD-free, propagation-invariant ST wave packet from Fig.~\ref{Fig:Measurements_k2}(a). Here we plot the initial wave packet at $z\!=\!0$ (dashed curve) in addition to the wave packet at $z\!=\!55$~mm (solid curve). The two profiles (initial and final) are practically identical. The initial temporal profile in panels (b) through (g) are all similar to the one plotted here. (b) The superluminal dispersive ST wave packets from Fig.~\ref{Fig:Measurements_k2}(b,c) with normal and anomalous second-order GVD; (c) the subluminal dispersive ST wave packets from Fig.~\ref{Fig:Measurements_k2}(e,f) with normal and anomalous second-order GVD; (d) from Fig.~\ref{Fig:Measurements_higherOrder}(a,b) for positive and negative third-order dispersion; (e) from Fig.~\ref{Fig:Measurements_higherOrder}(c,d) for positive and negative fourth-order dispersion; (f) from Fig.~\ref{Fig:Measurements_higherOrder}(e) and (g) from Fig.~\ref{Fig:Measurements_higherOrder}(f) for complex dispersion profiles. The top row depicts the dispersion orders implemented in each ST wave packet, each signified by an arrow if the magnitude is non-zero, with the direction of the arrow indicating the sign (the magnitudes themselves are not represented). In (b-e), the solid curve (dotted) curves correspond to the dispersion profile identified by solid (dotted) arrows. Note that we plot the profile along $t$ rather than the delayed frame $\tau$ in order to show the group delay incurred, which is related to the group velocity $\widetilde{v}\!=\!c\tan{\theta}$ in each case.}
\label{Fig:Measurements_temporal}
\end{figure*}

We repeat the measurements shown in Fig.~\ref{Fig:Measurements_k2}(a-c) carried out in the superluminal regime for ST wave packets in the subluminal regime  corresponding to $\theta\!=\!40^{\circ}$ ($\widetilde{v}\!\approx\!0.84c$) while retaining the same values for $k_{2}$, as shown in Fig.~\ref{Fig:Measurements_k2}(d-f). This confirms that the coefficients of the first two dispersion orders ($k_{1}$ and $k_{2}$ for GVD) are addressable \textit{independently} of each other. Once again, we note the different directions of pulse broadening [Fig.~\ref{Fig:Measurements_k2}(e) and Fig.~\ref{Fig:Measurements_k2}(f)] with respect to the propagation-invariant reference [Fig.~\ref{Fig:Measurements_k2}(d)]. Furthermore, note that the change in the direction of broadening changes between the superluminal and subluminal cases, as expected from theoretical calculations; see Fig.~\ref{Fig:Measurements_temporal} below.

We move on to demonstrating more complex dispersion profiles involving higher-order terms in $f(\Omega)$. First, we plot in Fig.~\ref{Fig:Measurements_higherOrder}(a,b) measurements for dispersive ST wave packets in which we isolate the third-order term in $f(\Omega)$, so that $k_{z}\!=\!k_{\mathrm{o}}+\Omega/\widetilde{v}+k_{3}\Omega^{3}/6$. Here we take $k_{3}\!=\!\pm80$~ps$^{3}$/mm. We also carry out measurements for dispersive ST wave packets in which we isolate the fourth-order term in $f(\Omega)$, so that $k_{z}\!=\!k_{\mathrm{o}}+\Omega/\widetilde{v}+k_{4}\Omega^{4}/24$; see Fig.~\ref{Fig:Measurements_higherOrder}(c,d). Here we take $k_{4}\!=\!\pm 40$~ps$^{4}$/mm. Finally, we produce two dispersive ST wave packets in which we select several higher-order dispersive terms to be realized simultaneously. In the example shown in Fig.~\ref{Fig:Measurements_higherOrder}(e), we implement a dispersion profile of the form $k_{z}\!=\!k_{\mathrm{o}}+\Omega/\widetilde{v}+|k_{2}|\Omega^{2}/2-|k_{3}|\Omega^{3}/6+|k_{4}|\Omega^{4}/24$. The dispersion profile in Fig.~\ref{Fig:Measurements_higherOrder}(f) has the same structure and the same value for $k_{2}$ and $k_{3}$, but we change the group velocity $\widetilde{v}$ and the fourth-order dispersion coefficient $k_{4}$, which leads to a major change in the overall dispersion profile, angular dispersion, and spatio-temporal profile evolution. In all cases, the measured angular dispersion is in excellent agreement with the theoretical expectations.

We summarize our results in Fig.~\ref{Fig:Measurements_temporal} where we plot the measured and theoretically predicted on-axis temporal profiles $I(0,z;\tau)$ at $z\!=\!55$~mm for the wave packets shown in Fig.~\ref{Fig:Measurements_k2} and Fig.~\ref{Fig:Measurements_higherOrder}. Note that the latter two cases [Fig.~\ref{Fig:Measurements_temporal}(f,g)] yield Airy-like pulse structures.

\section{Constraints on realizable dispersion profiles}

In principle, there are no fundamental constraints on the dispersion profile that can be realized employing the above-described strategy. Nevertheless, there are practical limitations due to unavoidably finite experimental resources. The crucial practical limit arises from the so-called `spectral uncertainty', which refers to the finite, albeit narrow, spectral bandwidth $\delta\lambda$ associated with each spatial frequency \cite{Yessenov19OE} rather than a single temporal frequency as implied in Eq.~\ref{Eq:SinPhiForGVDFree} and Eq.~\ref{Eq:GeneralAngleForDispersiveSTWP} -- not to be confused with the full spectral bandwidth $\Delta\lambda$; see Fig.~\ref{Fig:Measurements_k2}(a). The spectral uncertainty $\delta\lambda$ is observable in the finite width of the measured spectral projections in Fig.~\ref{Fig:Measurements_k2} and Fig.~\ref{Fig:Measurements_higherOrder}. This spectral uncertainty sets the upper limit on the propagation distance for ST wave packets, as well as on the maximum achievable differential group delay with respect to a luminal wave packet \cite{Yessenov19OE,Yessenov20NC}. From the perspective of the realizability of arbitrary GVD, $\delta\lambda$ determines the minimum distance between different states of dispersion that can be unambiguously implemented. Basically, two different but contiguous dispersion profiles are indistinguishable if there spectral projections overlap within $\delta\lambda$. In our setup as shown in Fig.~\ref{Fig:Setup}, $\delta\lambda$ is determined by the spectral resolution of the diffraction grating, and is estimated to be $\delta\lambda\!\sim\!25$~pm. Increasing the width of the grating improves its spectral resolution and reduces $\delta\lambda$, thereby allowing for more dispersion profiles to be distinguishable.

\begin{figure}[t!]
\centering
\includegraphics[width=8.6cm]{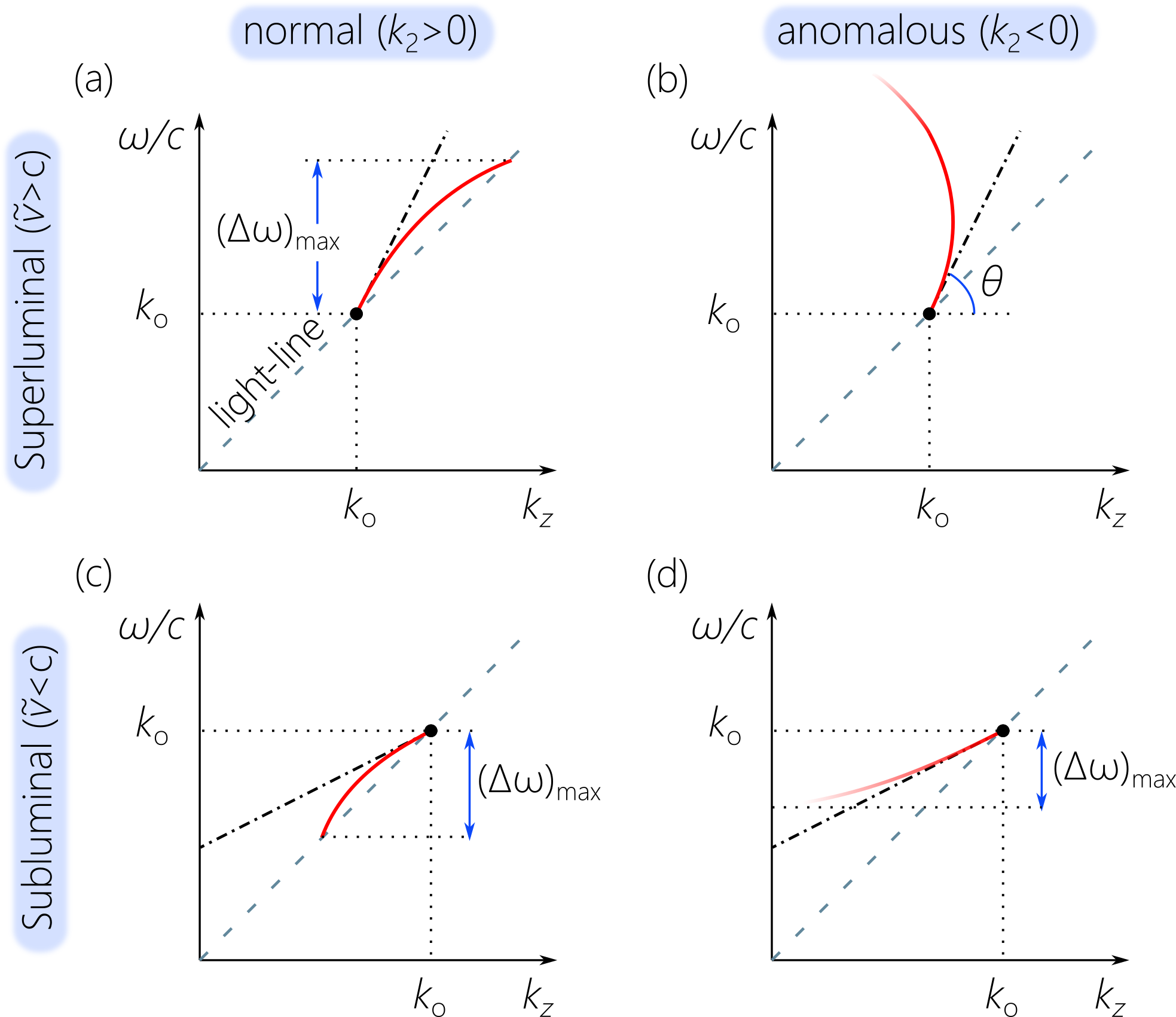}
\caption{Bandwidth limits over which control can be exercised over the GVD of ST wave packets. (a) Normal and (b) anomalous GVD in superluminal ST wave packets. (c,d) Same as (a,b) for subluminal ST wave packets. The solid red curves are the dispersion profiles, the dashed straight lines are the light-line $k_{z}\!=\!\omega/c$, and the dashed-dotted lines are the GVD-free projections $k_{z}\!=k_{\mathrm{o}}+(\omega-\omega_{\mathrm{o}})/\widetilde{v}$.}
\label{Fig:Constraints}
\end{figure}

Another constraint is associated with the maximum bandwidth over which dispersion control can be exercised. In general, the curved spectral trajectory for a ST wave packet projected onto the $(k_{z},\tfrac{\omega}{c})$-plane [Fig.~\ref{Fig:STWavePacketsAndAngularDispersion}(b), Fig.~\ref{Fig:Concept}, and Fig.~\ref{Fig:SecondOrderGVD}(i)] starts from the point $(k_{x},k_{z},\tfrac{\omega}{c})\!=\!(0,k_{\mathrm{o}},k_{\mathrm{o}})$ on the light-line. Deviation away from the light-line is associated with an increased spatial frequency $k_{x}$. The curved spectral projection in some cases may curve back and indeed re-intersect with the light-line, beyond which the field is evanescent. We consider here explicitly the case of second-order GVD $k_{z}\!=\!k_{\mathrm{o}}+\Omega/\widetilde{v}+\tfrac{1}{2}k_{2}\Omega^{2}$; a similar analysis can be applied to other dispersion profiles. In the \textit{superluminal} regime, we have $\omega\!>\!\omega_{\mathrm{o}}$, $\tan{\theta}\!>\!1$, and $\widetilde{v}\!>\!c$. For normal GVD $k_{2}\!>\!0$, the bandwidth is limited by the re-intersection of the curved spectral projection with the light-line to [Fig.~\ref{Fig:Constraints}(a)]:
\begin{equation}
(\Delta\omega)_{\mathrm{max}}=\omega_{\mathrm{max}}-\omega_{\mathrm{o}}=\frac{2}{ck_{2}}(1-\cot{\theta}).
\end{equation}
For anomalous GVD $k_{2}\!<\!0$, the bandwidth does not have a similar limit [Fig.~\ref{Fig:Constraints}(b)]. In the \textit{subluminal} regime, we have $\omega\!<\!\omega_{\mathrm{o}}$, $\tan{\theta}\!<\!1$, and $\widetilde{v}\!<\!c$. For normal GVD $k_{2}\!>\!0$, the bandwidth is once again limited by re-intersection of the dispersion profile with the light-line such that [Fig.~\ref{Fig:Constraints}(c)]:
\begin{equation}
(\Delta\omega)_{\mathrm{max}}=\frac{2}{ck_{2}}(\cot{\theta}-1).
\end{equation}
For anomalous GVD $k_{2}\!<\!0$, the bandwidth does not have a similar limit [Fig.~\ref{Fig:Constraints}(d)].

More generally, for a given system numerical aperture (NA) within the paraxial regime, the bandwidth must satisfy the following inequality:
\begin{equation}
(\mathrm{NA})^{2}>\eta^{2}\,\,\frac{\Delta\omega}{\omega_{\mathrm{o}}}\left(1+\frac{1+\widetilde{n}}{2}\cdot\frac{\Delta\omega}{\omega_{\mathrm{o}}}\right).
\end{equation}
This criterion provides an estimate of the accessible bandwidth for any particular dispersion profile. Of course, increasing the bandwidth can extend the dispersion profile until it potentially intersects with the $\omega$-axis ($k_{z}\!=\!0$); negative-valued $k_{z}$ beyond this limit are excluded because they are incompatible with causal excitation and propagation \cite{Yessenov19PRA}. However, before reaching this limit, the field becomes non-paraxial ($k_{x}\!\sim\!k_{\mathrm{o}}$) and an altogether different analysis is required. 

\section{Discussion and Conclusions}

To date, most work on ST wave packets in free space has focused to date on propagation invariance, whether for X-waves \cite{Saari97PRL}, sideband ST wave packets such as focus-wave modes \cite{Brittingham83JAP,Reivelt00JOSAA,Reivelt02PRE}, or baseband ST wave packets \cite{Kondakci16OE,Parker16OE,Kondakci17NP,Bhaduri18OE,Bhaduri19OL} (according to the classification in \cite{Yessenov19PRA}). Our work reported here is part of an ongoing effort regarding the control over the axial evolution of ST wave packets, which encompasses axial spectral encoding \cite{Motz20arxiv} (producing a prescribed axial evolution of the pulse spectrum along the propagation axis), and accelerating or decelerating wave packets whose group velocity varies axially \cite{Yessenov20PRL2}. Here, the temporal profile of the freely propagating ST wave packet is dictated by incorporating an effective GVD in free space.

Previous efforts directed at ST wave packets \textit{in} dispersive media \cite{Longhi04OL,Porras03PRE2,Porras05OL,Malaguti08OL,Malaguti09PRA} study their properties within these media, and are thus appropriate for those cases where the ST wave packet is  generated there via nonlinear optical effects. There have been no examples of launching a ST wave packet rationally synthesized in free space into the medium to achieve a desired effect (with the exception of the early experiment in \cite{Sonajalg97OL} on X-waves). The results presented here, in addition to our recent work on formulating laws of refraction for ST wave packets across planar interfaces \cite{Bhaduri20NP}, therefore lay the experimental foundations for this endeavor.

Our results give rise to a fundamental question: why do TPFs offer only limited controllability over the GVD experienced upon free propagation, whereas ST wave packets offer the potential for arbitrary control over the dispersion profile, despite the fact that both TPFs and ST wave packets are undergirded by angular dispersion? As we have shown, the critical distinction is realted to the differentiability of the angular dispersion associated with these two field configurations: the angular dispersion underlying TPFs is differentiable, whereas that for ST wave packets is non-differentiable. The surprising impact of the abstract mathematical notion of differentiability on dispersion control can be justified as follows: the differentiability of the angular dispersion underlying TPFs indicates that only a few parameters in the expansion in Eq.~\ref{Eq:ExpansionForPhi} are accessible, especially for narrow bandwidths. This is inescapable in light of the differentiability of the angular dispersion $\varphi(\omega)$. On the other hand, the non-differentiability of $\varphi(\omega)$ for ST wave packets indicates the availability of many more degrees of freedom, which may become accessible even within narrow bandwidths. 

In conclusion, we have demonstrated theoretically and experimentally that ST wave packets can be endowed with arbitrary dispersion profiles in free space through sculpting their angular dispersion. In contrast to conventional angular dispersion which restricts the realizable GVD to the anomalous regime, ST wave packets can be designed to experience anomalous or normal GVD. In fact, the dispersion profile of ST wave packets can be moulded to an arbitrary profile, whereby the amplitude \textit{and} sign of each dispersion order can be addressed separately, independently of the others, and set to values unattainable in traditional optical materials far from resonance -- while not introducing optical losses. The unique feature of ST wave packets that enables such versatility is the non-differentiability of their underlying angular dispersion, in contrast to the differentiability of that associated with TPFs and other previously studied wave packets (including X-waves and pulsed Bessel beams). This strategy can be viewed as engineering the optical vacuum: enabling the inculcation of arbitrary dispersion profiles (including the magnitude and sign of all dispersion orders) in freely propagating structured fields. Our results reported here represent a critical stepping stone towards the study of the propagation of ST wave packets in dispersive media, whereby the internal spatio-temporal structure of the wave packets can be exploited to circumvent traditional material constraints and thus help explore new phenomena in nonlinear and quantum optics.

\section*{Acknowledgments}
This work was supported by the U.S. Office of Naval Research
(ONR) under Contract N00014-17-1-2458, and ONR MURI
contract N00014-20-1-2789.

\bibliography{diffraction}

\begin{thebibliography}{82}%
\makeatletter
\providecommand \@ifxundefined [1]{%
 \@ifx{#1\undefined}
}%
\providecommand \@ifnum [1]{%
 \ifnum #1\expandafter \@firstoftwo
 \else \expandafter \@secondoftwo
 \fi
}%
\providecommand \@ifx [1]{%
 \ifx #1\expandafter \@firstoftwo
 \else \expandafter \@secondoftwo
 \fi
}%
\providecommand \natexlab [1]{#1}%
\providecommand \enquote  [1]{``#1''}%
\providecommand \bibnamefont  [1]{#1}%
\providecommand \bibfnamefont [1]{#1}%
\providecommand \citenamefont [1]{#1}%
\providecommand \href@noop [0]{\@secondoftwo}%
\providecommand \href [0]{\begingroup \@sanitize@url \@href}%
\providecommand \@href[1]{\@@startlink{#1}\@@href}%
\providecommand \@@href[1]{\endgroup#1\@@endlink}%
\providecommand \@sanitize@url [0]{\catcode `\\12\catcode `\$12\catcode
  `\&12\catcode `\#12\catcode `\^12\catcode `\_12\catcode `\%12\relax}%
\providecommand \@@startlink[1]{}%
\providecommand \@@endlink[0]{}%
\providecommand \url  [0]{\begingroup\@sanitize@url \@url }%
\providecommand \@url [1]{\endgroup\@href {#1}{\urlprefix }}%
\providecommand \urlprefix  [0]{URL }%
\providecommand \Eprint [0]{\href }%
\providecommand \doibase [0]{http://dx.doi.org/}%
\providecommand \selectlanguage [0]{\@gobble}%
\providecommand \bibinfo  [0]{\@secondoftwo}%
\providecommand \bibfield  [0]{\@secondoftwo}%
\providecommand \translation [1]{[#1]}%
\providecommand \BibitemOpen [0]{}%
\providecommand \bibitemStop [0]{}%
\providecommand \bibitemNoStop [0]{.\EOS\space}%
\providecommand \EOS [0]{\spacefactor3000\relax}%
\providecommand \BibitemShut  [1]{\csname bibitem#1\endcsname}%
\let\auto@bib@innerbib\@empty
\bibitem [{\citenamefont {Saleh}\ and\ \citenamefont
  {Teich}(2007)}]{SalehBook07}%
  \BibitemOpen
  \bibfield  {author} {\bibinfo {author} {\bibfnamefont {B.~E.~A.}\
  \bibnamefont {Saleh}}\ and\ \bibinfo {author} {\bibfnamefont {M.~C.}\
  \bibnamefont {Teich}},\ }\href@noop {} {\emph {\bibinfo {title} {Principles
  of Photonics}}}\ (\bibinfo  {publisher} {Wiley},\ \bibinfo {year}
  {2007})\BibitemShut {NoStop}%
\bibitem [{\citenamefont {Akturk}\ \emph {et~al.}(2005)\citenamefont {Akturk},
  \citenamefont {Gu}, \citenamefont {Gabolde},\ and\ \citenamefont
  {Trebino}}]{Akturk05OE}%
  \BibitemOpen
  \bibfield  {author} {\bibinfo {author} {\bibfnamefont {S.}~\bibnamefont
  {Akturk}}, \bibinfo {author} {\bibfnamefont {X.}~\bibnamefont {Gu}}, \bibinfo
  {author} {\bibfnamefont {P.}~\bibnamefont {Gabolde}}, \ and\ \bibinfo
  {author} {\bibfnamefont {R.}~\bibnamefont {Trebino}},\ }\href@noop {}
  {\bibfield  {journal} {\bibinfo  {journal} {Opt. Express}\ }\textbf {\bibinfo
  {volume} {13}},\ \bibinfo {pages} {8642} (\bibinfo {year}
  {2005})}\BibitemShut {NoStop}%
\bibitem [{\citenamefont {Dorrer}(2019)}]{Dorrer19IEEE}%
  \BibitemOpen
  \bibfield  {author} {\bibinfo {author} {\bibfnamefont {C.}~\bibnamefont
  {Dorrer}},\ }\href@noop {} {\bibfield  {journal} {\bibinfo  {journal} {IEEE
  J. Sel. Top. Quantum Electron.}\ }\textbf {\bibinfo {volume} {25}},\ \bibinfo
  {pages} {3100216} (\bibinfo {year} {2019})}\BibitemShut {NoStop}%
\bibitem [{\citenamefont {Giovannini}\ \emph {et~al.}(2015)\citenamefont
  {Giovannini}, \citenamefont {Romero}, \citenamefont {Poto{\v c}},
  \citenamefont {Ferenczi}, \citenamefont {Speirits}, \citenamefont {Barnett},
  \citenamefont {Faccio},\ and\ \citenamefont {Padgett}}]{Giovannini15Science}%
  \BibitemOpen
  \bibfield  {author} {\bibinfo {author} {\bibfnamefont {D.}~\bibnamefont
  {Giovannini}}, \bibinfo {author} {\bibfnamefont {J.}~\bibnamefont {Romero}},
  \bibinfo {author} {\bibfnamefont {V.}~\bibnamefont {Poto{\v c}}}, \bibinfo
  {author} {\bibfnamefont {G.}~\bibnamefont {Ferenczi}}, \bibinfo {author}
  {\bibfnamefont {F.}~\bibnamefont {Speirits}}, \bibinfo {author}
  {\bibfnamefont {S.~M.}\ \bibnamefont {Barnett}}, \bibinfo {author}
  {\bibfnamefont {D.}~\bibnamefont {Faccio}}, \ and\ \bibinfo {author}
  {\bibfnamefont {M.~J.}\ \bibnamefont {Padgett}},\ }\href@noop {} {\bibfield
  {journal} {\bibinfo  {journal} {Science}\ }\textbf {\bibinfo {volume}
  {347}},\ \bibinfo {pages} {857} (\bibinfo {year} {2015})}\BibitemShut
  {NoStop}%
\bibitem [{\citenamefont {Alfano}\ and\ \citenamefont
  {Nolan}(2016)}]{Alfano16OC}%
  \BibitemOpen
  \bibfield  {author} {\bibinfo {author} {\bibfnamefont {R.~R.}\ \bibnamefont
  {Alfano}}\ and\ \bibinfo {author} {\bibfnamefont {D.~A.}\ \bibnamefont
  {Nolan}},\ }\href@noop {} {\bibfield  {journal} {\bibinfo  {journal} {Opt.
  Commun.}\ }\textbf {\bibinfo {volume} {361}},\ \bibinfo {pages} {25}
  (\bibinfo {year} {2016})}\BibitemShut {NoStop}%
\bibitem [{\citenamefont {Bouchard}\ \emph {et~al.}(2016)\citenamefont
  {Bouchard}, \citenamefont {Harris}, \citenamefont {Mand}, \citenamefont
  {Boyd},\ and\ \citenamefont {Karimi}}]{Bouchard16Optica}%
  \BibitemOpen
  \bibfield  {author} {\bibinfo {author} {\bibfnamefont {F.}~\bibnamefont
  {Bouchard}}, \bibinfo {author} {\bibfnamefont {J.}~\bibnamefont {Harris}},
  \bibinfo {author} {\bibfnamefont {H.}~\bibnamefont {Mand}}, \bibinfo {author}
  {\bibfnamefont {R.~W.}\ \bibnamefont {Boyd}}, \ and\ \bibinfo {author}
  {\bibfnamefont {E.}~\bibnamefont {Karimi}},\ }\href@noop {} {\bibfield
  {journal} {\bibinfo  {journal} {Optica}\ }\textbf {\bibinfo {volume} {3}},\
  \bibinfo {pages} {351} (\bibinfo {year} {2016})}\BibitemShut {NoStop}%
\bibitem [{\citenamefont {Lyons}\ \emph {et~al.}(2018)\citenamefont {Lyons},
  \citenamefont {Roger}, \citenamefont {Westerberg}, \citenamefont {Vezzoli},
  \citenamefont {Maitland}, \citenamefont {Leach}, \citenamefont {Padgett},\
  and\ \citenamefont {Faccio}}]{Lyons18Optica}%
  \BibitemOpen
  \bibfield  {author} {\bibinfo {author} {\bibfnamefont {A.}~\bibnamefont
  {Lyons}}, \bibinfo {author} {\bibfnamefont {T.}~\bibnamefont {Roger}},
  \bibinfo {author} {\bibfnamefont {N.}~\bibnamefont {Westerberg}}, \bibinfo
  {author} {\bibfnamefont {S.}~\bibnamefont {Vezzoli}}, \bibinfo {author}
  {\bibfnamefont {C.}~\bibnamefont {Maitland}}, \bibinfo {author}
  {\bibfnamefont {J.}~\bibnamefont {Leach}}, \bibinfo {author} {\bibfnamefont
  {M.~J.}\ \bibnamefont {Padgett}}, \ and\ \bibinfo {author} {\bibfnamefont
  {D.}~\bibnamefont {Faccio}},\ }\href@noop {} {\bibfield  {journal} {\bibinfo
  {journal} {Optica}\ }\textbf {\bibinfo {volume} {5}},\ \bibinfo {pages} {682}
  (\bibinfo {year} {2018})}\BibitemShut {NoStop}%
\bibitem [{\citenamefont {Zapata-Rodr{\'i}guez}\ \emph
  {et~al.}(2008)\citenamefont {Zapata-Rodr{\'i}guez}, \citenamefont {Porras},\
  and\ \citenamefont {Miret}}]{Zapata08JOSAA}%
  \BibitemOpen
  \bibfield  {author} {\bibinfo {author} {\bibfnamefont {C.~J.}\ \bibnamefont
  {Zapata-Rodr{\'i}guez}}, \bibinfo {author} {\bibfnamefont {M.~A.}\
  \bibnamefont {Porras}}, \ and\ \bibinfo {author} {\bibfnamefont {J.~J.}\
  \bibnamefont {Miret}},\ }\href@noop {} {\bibfield  {journal} {\bibinfo
  {journal} {J. Opt. Soc. Am. A}\ }\textbf {\bibinfo {volume} {25}},\ \bibinfo
  {pages} {2758} (\bibinfo {year} {2008})}\BibitemShut {NoStop}%
\bibitem [{\citenamefont {Saari}(2017)}]{Saari17OC}%
  \BibitemOpen
  \bibfield  {author} {\bibinfo {author} {\bibfnamefont {P.}~\bibnamefont
  {Saari}},\ }\href@noop {} {\bibfield  {journal} {\bibinfo  {journal} {Opt.
  Commun.}\ }\textbf {\bibinfo {volume} {392}},\ \bibinfo {pages} {300}
  (\bibinfo {year} {2017})}\BibitemShut {NoStop}%
\bibitem [{\citenamefont {Zhu}\ \emph {et~al.}(2005)\citenamefont {Zhu},
  \citenamefont {{van H}owe}, \citenamefont {Durst}, \citenamefont {Zipfel},\
  and\ \citenamefont {Xu}}]{Zhu05OE}%
  \BibitemOpen
  \bibfield  {author} {\bibinfo {author} {\bibfnamefont {G.}~\bibnamefont
  {Zhu}}, \bibinfo {author} {\bibfnamefont {J.}~\bibnamefont {{van H}owe}},
  \bibinfo {author} {\bibfnamefont {M.}~\bibnamefont {Durst}}, \bibinfo
  {author} {\bibfnamefont {W.}~\bibnamefont {Zipfel}}, \ and\ \bibinfo {author}
  {\bibfnamefont {C.}~\bibnamefont {Xu}},\ }\href@noop {} {\bibfield  {journal}
  {\bibinfo  {journal} {Opt. Express}\ }\textbf {\bibinfo {volume} {13}},\
  \bibinfo {pages} {2153} (\bibinfo {year} {2005})}\BibitemShut {NoStop}%
\bibitem [{\citenamefont {{Di T}rapani}\ \emph {et~al.}(1998)\citenamefont {{Di
  T}rapani}, \citenamefont {D}, \citenamefont {Valiulis}, \citenamefont
  {Dubietis}, \citenamefont {Danielius},\ and\ \citenamefont
  {Piskarskas}}]{DiTrapani98PRL}%
  \BibitemOpen
  \bibfield  {author} {\bibinfo {author} {\bibfnamefont {P.}~\bibnamefont {{Di
  T}rapani}}, \bibinfo {author} {\bibfnamefont {C.}~\bibnamefont {D}}, \bibinfo
  {author} {\bibfnamefont {G.}~\bibnamefont {Valiulis}}, \bibinfo {author}
  {\bibfnamefont {A.}~\bibnamefont {Dubietis}}, \bibinfo {author}
  {\bibfnamefont {R.}~\bibnamefont {Danielius}}, \ and\ \bibinfo {author}
  {\bibfnamefont {A.}~\bibnamefont {Piskarskas}},\ }\href@noop {} {\bibfield
  {journal} {\bibinfo  {journal} {Phys. Rev. Lett.}\ }\textbf {\bibinfo
  {volume} {82}},\ \bibinfo {pages} {570} (\bibinfo {year} {1998})}\BibitemShut
  {NoStop}%
\bibitem [{\citenamefont {Porras}\ \emph {et~al.}(2005)\citenamefont {Porras},
  \citenamefont {Dubietis}, \citenamefont {Ku{\u c}inskas}, \citenamefont
  {Bragheri}, \citenamefont {Degiorgio}, \citenamefont {Couairon},
  \citenamefont {Faccio},\ and\ \citenamefont {{Di T}rapani}}]{Porras05OL}%
  \BibitemOpen
  \bibfield  {author} {\bibinfo {author} {\bibfnamefont {M.~A.}\ \bibnamefont
  {Porras}}, \bibinfo {author} {\bibfnamefont {A.}~\bibnamefont {Dubietis}},
  \bibinfo {author} {\bibfnamefont {E.}~\bibnamefont {Ku{\u c}inskas}},
  \bibinfo {author} {\bibfnamefont {F.}~\bibnamefont {Bragheri}}, \bibinfo
  {author} {\bibfnamefont {V.}~\bibnamefont {Degiorgio}}, \bibinfo {author}
  {\bibfnamefont {A.}~\bibnamefont {Couairon}}, \bibinfo {author}
  {\bibfnamefont {D.}~\bibnamefont {Faccio}}, \ and\ \bibinfo {author}
  {\bibfnamefont {P.}~\bibnamefont {{Di T}rapani}},\ }\href@noop {} {\bibfield
  {journal} {\bibinfo  {journal} {Opt. Lett.}\ }\textbf {\bibinfo {volume}
  {30}},\ \bibinfo {pages} {3398} (\bibinfo {year} {2005})}\BibitemShut
  {NoStop}%
\bibitem [{\citenamefont {Faccio}\ \emph {et~al.}(2007)\citenamefont {Faccio},
  \citenamefont {Averchi}, \citenamefont {Couairon}, \citenamefont {Kolesik},
  \citenamefont {Moloney}, \citenamefont {Dubietis}, \citenamefont
  {Tamosauskas}, \citenamefont {Polesana}, \citenamefont {Piskarskas},\ and\
  \citenamefont {Trapani}}]{Faccio07OE}%
  \BibitemOpen
  \bibfield  {author} {\bibinfo {author} {\bibfnamefont {D.}~\bibnamefont
  {Faccio}}, \bibinfo {author} {\bibfnamefont {A.}~\bibnamefont {Averchi}},
  \bibinfo {author} {\bibfnamefont {A.}~\bibnamefont {Couairon}}, \bibinfo
  {author} {\bibfnamefont {M.}~\bibnamefont {Kolesik}}, \bibinfo {author}
  {\bibfnamefont {J.}~\bibnamefont {Moloney}}, \bibinfo {author} {\bibfnamefont
  {A.}~\bibnamefont {Dubietis}}, \bibinfo {author} {\bibfnamefont
  {G.}~\bibnamefont {Tamosauskas}}, \bibinfo {author} {\bibfnamefont
  {P.}~\bibnamefont {Polesana}}, \bibinfo {author} {\bibfnamefont
  {A.}~\bibnamefont {Piskarskas}}, \ and\ \bibinfo {author} {\bibfnamefont
  {P.~D.}\ \bibnamefont {Trapani}},\ }\href@noop {} {\bibfield  {journal}
  {\bibinfo  {journal} {Opt. Express}\ }\textbf {\bibinfo {volume} {15}},\
  \bibinfo {pages} {13077} (\bibinfo {year} {2007})}\BibitemShut {NoStop}%
\bibitem [{\citenamefont {Malaguti}\ \emph {et~al.}(2008)\citenamefont
  {Malaguti}, \citenamefont {Bellanca},\ and\ \citenamefont
  {Trillo}}]{Malaguti08OL}%
  \BibitemOpen
  \bibfield  {author} {\bibinfo {author} {\bibfnamefont {S.}~\bibnamefont
  {Malaguti}}, \bibinfo {author} {\bibfnamefont {G.}~\bibnamefont {Bellanca}},
  \ and\ \bibinfo {author} {\bibfnamefont {S.}~\bibnamefont {Trillo}},\
  }\href@noop {} {\bibfield  {journal} {\bibinfo  {journal} {Opt. Lett.}\
  }\textbf {\bibinfo {volume} {33}},\ \bibinfo {pages} {1117} (\bibinfo {year}
  {2008})}\BibitemShut {NoStop}%
\bibitem [{\citenamefont {Malaguti}\ and\ \citenamefont
  {Trillo}(2009)}]{Malaguti09PRA}%
  \BibitemOpen
  \bibfield  {author} {\bibinfo {author} {\bibfnamefont {S.}~\bibnamefont
  {Malaguti}}\ and\ \bibinfo {author} {\bibfnamefont {S.}~\bibnamefont
  {Trillo}},\ }\href@noop {} {\bibfield  {journal} {\bibinfo  {journal} {Phys.
  Rev. A}\ }\textbf {\bibinfo {volume} {79}},\ \bibinfo {pages} {063803}
  (\bibinfo {year} {2009})}\BibitemShut {NoStop}%
\bibitem [{\citenamefont {Katamadze}\ \emph {et~al.}(2015)\citenamefont
  {Katamadze}, \citenamefont {Borshchevskaya}, \citenamefont {Dyakonov},
  \citenamefont {Paterova},\ and\ \citenamefont {Kulik}}]{Katamadze15PRA}%
  \BibitemOpen
  \bibfield  {author} {\bibinfo {author} {\bibfnamefont {K.~G.}\ \bibnamefont
  {Katamadze}}, \bibinfo {author} {\bibfnamefont {N.~A.}\ \bibnamefont
  {Borshchevskaya}}, \bibinfo {author} {\bibfnamefont {I.~V.}\ \bibnamefont
  {Dyakonov}}, \bibinfo {author} {\bibfnamefont {A.~V.}\ \bibnamefont
  {Paterova}}, \ and\ \bibinfo {author} {\bibfnamefont {S.~P.}\ \bibnamefont
  {Kulik}},\ }\href@noop {} {\bibfield  {journal} {\bibinfo  {journal} {Phys.
  Rev. A}\ }\textbf {\bibinfo {volume} {92}},\ \bibinfo {pages} {023812}
  (\bibinfo {year} {2015})}\BibitemShut {NoStop}%
\bibitem [{\citenamefont {Spasibko}\ \emph {et~al.}(2016)\citenamefont
  {Spasibko}, \citenamefont {Kopylov}, \citenamefont {Murzina}, \citenamefont
  {Leuchs},\ and\ \citenamefont {Chekhova}}]{Spasibko16OL}%
  \BibitemOpen
  \bibfield  {author} {\bibinfo {author} {\bibfnamefont {K.~Y.}\ \bibnamefont
  {Spasibko}}, \bibinfo {author} {\bibfnamefont {D.~A.}\ \bibnamefont
  {Kopylov}}, \bibinfo {author} {\bibfnamefont {T.~V.}\ \bibnamefont
  {Murzina}}, \bibinfo {author} {\bibfnamefont {G.}~\bibnamefont {Leuchs}}, \
  and\ \bibinfo {author} {\bibfnamefont {M.~V.}\ \bibnamefont {Chekhova}},\
  }\href@noop {} {\bibfield  {journal} {\bibinfo  {journal} {Opt. Lett.}\
  }\textbf {\bibinfo {volume} {41}},\ \bibinfo {pages} {2827} (\bibinfo {year}
  {2016})}\BibitemShut {NoStop}%
\bibitem [{\citenamefont {Cutipa}\ \emph {et~al.}(2020)\citenamefont {Cutipa},
  \citenamefont {Spasibko},\ and\ \citenamefont {Chekhova}}]{Cupita20OL}%
  \BibitemOpen
  \bibfield  {author} {\bibinfo {author} {\bibfnamefont {P.}~\bibnamefont
  {Cutipa}}, \bibinfo {author} {\bibfnamefont {K.~Y.}\ \bibnamefont
  {Spasibko}}, \ and\ \bibinfo {author} {\bibfnamefont {M.~V.}\ \bibnamefont
  {Chekhova}},\ }\href@noop {} {\bibfield  {journal} {\bibinfo  {journal} {Opt.
  Lett.}\ }\textbf {\bibinfo {volume} {45}},\ \bibinfo {pages} {3581} (\bibinfo
  {year} {2020})}\BibitemShut {NoStop}%
\bibitem [{\citenamefont {Torres}\ \emph {et~al.}(2010)\citenamefont {Torres},
  \citenamefont {Hendrych},\ and\ \citenamefont {Valencia}}]{Torres10AOP}%
  \BibitemOpen
  \bibfield  {author} {\bibinfo {author} {\bibfnamefont {J.~P.}\ \bibnamefont
  {Torres}}, \bibinfo {author} {\bibfnamefont {M.}~\bibnamefont {Hendrych}}, \
  and\ \bibinfo {author} {\bibfnamefont {A.}~\bibnamefont {Valencia}},\
  }\href@noop {} {\bibfield  {journal} {\bibinfo  {journal} {Adv. Opt.
  Photon.}\ }\textbf {\bibinfo {volume} {2}},\ \bibinfo {pages} {319} (\bibinfo
  {year} {2010})}\BibitemShut {NoStop}%
\bibitem [{\citenamefont {Bor}\ and\ \citenamefont {R{\'a}cz}(1985)}]{Bor85OC}%
  \BibitemOpen
  \bibfield  {author} {\bibinfo {author} {\bibfnamefont {Z.}~\bibnamefont
  {Bor}}\ and\ \bibinfo {author} {\bibfnamefont {B.}~\bibnamefont {R{\'a}cz}},\
  }\href@noop {} {\bibfield  {journal} {\bibinfo  {journal} {Opt. Commun.}\
  }\textbf {\bibinfo {volume} {54}},\ \bibinfo {pages} {165} (\bibinfo {year}
  {1985})}\BibitemShut {NoStop}%
\bibitem [{\citenamefont {Hebling}(1996)}]{Hebling96OQE}%
  \BibitemOpen
  \bibfield  {author} {\bibinfo {author} {\bibfnamefont {J.}~\bibnamefont
  {Hebling}},\ }\href@noop {} {\bibfield  {journal} {\bibinfo  {journal} {Opt.
  Quant. Electron.}\ }\textbf {\bibinfo {volume} {28}},\ \bibinfo {pages}
  {1759} (\bibinfo {year} {1996})}\BibitemShut {NoStop}%
\bibitem [{\citenamefont {F{\"u}l{\"o}p}\ and\ \citenamefont
  {Hebling}(2010)}]{Fulop10Review}%
  \BibitemOpen
  \bibfield  {author} {\bibinfo {author} {\bibfnamefont {J.~A.}\ \bibnamefont
  {F{\"u}l{\"o}p}}\ and\ \bibinfo {author} {\bibfnamefont {J.}~\bibnamefont
  {Hebling}},\ }in\ \href@noop {} {\emph {\bibinfo {booktitle} {Recent Optical
  and Photonic Technologies}}},\ \bibinfo {editor} {edited by\ \bibinfo
  {editor} {\bibfnamefont {K.~Y.}\ \bibnamefont {Kim}}}\ (\bibinfo  {publisher}
  {InTech},\ \bibinfo {year} {2010})\BibitemShut {NoStop}%
\bibitem [{\citenamefont {Martinez}\ \emph {et~al.}(1984)\citenamefont
  {Martinez}, \citenamefont {Gordon},\ and\ \citenamefont
  {Fork}}]{Martinez84JOSAA}%
  \BibitemOpen
  \bibfield  {author} {\bibinfo {author} {\bibfnamefont {O.~E.}\ \bibnamefont
  {Martinez}}, \bibinfo {author} {\bibfnamefont {J.~P.}\ \bibnamefont
  {Gordon}}, \ and\ \bibinfo {author} {\bibfnamefont {R.~L.}\ \bibnamefont
  {Fork}},\ }\href@noop {} {\bibfield  {journal} {\bibinfo  {journal} {J. Opt.
  Soc. Am. A}\ }\textbf {\bibinfo {volume} {1}},\ \bibinfo {pages} {1003}
  (\bibinfo {year} {1984})}\BibitemShut {NoStop}%
\bibitem [{\citenamefont {Porras}\ \emph {et~al.}(2003)\citenamefont {Porras},
  \citenamefont {Valiulis},\ and\ \citenamefont {{Di T}rapani}}]{Porras03PRE2}%
  \BibitemOpen
  \bibfield  {author} {\bibinfo {author} {\bibfnamefont {M.~A.}\ \bibnamefont
  {Porras}}, \bibinfo {author} {\bibfnamefont {G.}~\bibnamefont {Valiulis}}, \
  and\ \bibinfo {author} {\bibfnamefont {P.}~\bibnamefont {{Di T}rapani}},\
  }\href@noop {} {\bibfield  {journal} {\bibinfo  {journal} {Phys. Rev. E}\
  }\textbf {\bibinfo {volume} {68}},\ \bibinfo {pages} {016613} (\bibinfo
  {year} {2003})}\BibitemShut {NoStop}%
\bibitem [{\citenamefont {Martinez}(1989)}]{Martinez89IEEE}%
  \BibitemOpen
  \bibfield  {author} {\bibinfo {author} {\bibfnamefont {O.~E.}\ \bibnamefont
  {Martinez}},\ }\href@noop {} {\bibfield  {journal} {\bibinfo  {journal} {IEEE
  J. Sel. Top. Quantum Electron.}\ }\textbf {\bibinfo {volume} {25}},\ \bibinfo
  {pages} {2464} (\bibinfo {year} {1989})}\BibitemShut {NoStop}%
\bibitem [{\citenamefont {Szab{\'o}}\ and\ \citenamefont
  {Bor}(1990)}]{Szabo90APB}%
  \BibitemOpen
  \bibfield  {author} {\bibinfo {author} {\bibfnamefont {G.}~\bibnamefont
  {Szab{\'o}}}\ and\ \bibinfo {author} {\bibfnamefont {Z.}~\bibnamefont
  {Bor}},\ }\href@noop {} {\bibfield  {journal} {\bibinfo  {journal} {Appl.
  Phys. B}\ }\textbf {\bibinfo {volume} {58}},\ \bibinfo {pages} {51} (\bibinfo
  {year} {1990})}\BibitemShut {NoStop}%
\bibitem [{\citenamefont {Dubietis}\ \emph {et~al.}(1997)\citenamefont
  {Dubietis}, \citenamefont {Valiulis}, \citenamefont {Tamosauskas},
  \citenamefont {Danielius},\ and\ \citenamefont {Piskarskas}}]{Dubietis97OL}%
  \BibitemOpen
  \bibfield  {author} {\bibinfo {author} {\bibfnamefont {A.}~\bibnamefont
  {Dubietis}}, \bibinfo {author} {\bibfnamefont {G.}~\bibnamefont {Valiulis}},
  \bibinfo {author} {\bibfnamefont {G.}~\bibnamefont {Tamosauskas}}, \bibinfo
  {author} {\bibfnamefont {R.}~\bibnamefont {Danielius}}, \ and\ \bibinfo
  {author} {\bibfnamefont {A.}~\bibnamefont {Piskarskas}},\ }\href@noop {}
  {\bibfield  {journal} {\bibinfo  {journal} {Opt. Lett.}\ }\textbf {\bibinfo
  {volume} {22}},\ \bibinfo {pages} {1071} (\bibinfo {year}
  {1997})}\BibitemShut {NoStop}%
\bibitem [{\citenamefont {Liu}\ \emph {et~al.}(2000)\citenamefont {Liu},
  \citenamefont {Beckwitt},\ and\ \citenamefont {Wise}}]{Liu00PRE}%
  \BibitemOpen
  \bibfield  {author} {\bibinfo {author} {\bibfnamefont {X.}~\bibnamefont
  {Liu}}, \bibinfo {author} {\bibfnamefont {K.}~\bibnamefont {Beckwitt}}, \
  and\ \bibinfo {author} {\bibfnamefont {F.~W.}\ \bibnamefont {Wise}},\
  }\href@noop {} {\bibfield  {journal} {\bibinfo  {journal} {Phys. Rev. E}\
  }\textbf {\bibinfo {volume} {62}},\ \bibinfo {pages} {1328} (\bibinfo {year}
  {2000})}\BibitemShut {NoStop}%
\bibitem [{\citenamefont {Wise}\ and\ \citenamefont {{Di
  T}rapani}(2002)}]{Wise02OPN}%
  \BibitemOpen
  \bibfield  {author} {\bibinfo {author} {\bibfnamefont {F.}~\bibnamefont
  {Wise}}\ and\ \bibinfo {author} {\bibfnamefont {P.}~\bibnamefont {{Di
  T}rapani}},\ }\href@noop {} {\bibfield  {journal} {\bibinfo  {journal} {Opt.
  Photon. News}\ }\textbf {\bibinfo {volume} {13}},\ \bibinfo {pages} {29}
  (\bibinfo {year} {2002})}\BibitemShut {NoStop}%
\bibitem [{\citenamefont {Schober}\ \emph {et~al.}(2007)\citenamefont
  {Schober}, \citenamefont {Charbonneau-Lefort},\ and\ \citenamefont
  {Fejer}}]{Schober07JOSAB}%
  \BibitemOpen
  \bibfield  {author} {\bibinfo {author} {\bibfnamefont {A.}~\bibnamefont
  {Schober}}, \bibinfo {author} {\bibfnamefont {M.}~\bibnamefont
  {Charbonneau-Lefort}}, \ and\ \bibinfo {author} {\bibfnamefont
  {M.}~\bibnamefont {Fejer}},\ }\href@noop {} {\bibfield  {journal} {\bibinfo
  {journal} {J. Opt. Soc. Am. B}\ }\textbf {\bibinfo {volume} {22}},\ \bibinfo
  {pages} {1699} (\bibinfo {year} {2007})}\BibitemShut {NoStop}%
\bibitem [{\citenamefont {Torres}\ \emph {et~al.}(2005)\citenamefont {Torres},
  \citenamefont {Mitchell},\ and\ \citenamefont {Hendrych}}]{Torres05PRA}%
  \BibitemOpen
  \bibfield  {author} {\bibinfo {author} {\bibfnamefont {J.~P.}\ \bibnamefont
  {Torres}}, \bibinfo {author} {\bibfnamefont {M.~W.}\ \bibnamefont
  {Mitchell}}, \ and\ \bibinfo {author} {\bibfnamefont {M.}~\bibnamefont
  {Hendrych}},\ }\href@noop {} {\bibfield  {journal} {\bibinfo  {journal}
  {Phys. Rev. A}\ }\textbf {\bibinfo {volume} {71}},\ \bibinfo {pages} {022320}
  (\bibinfo {year} {2005})}\BibitemShut {NoStop}%
\bibitem [{\citenamefont {Hendrych}\ \emph {et~al.}(2009)\citenamefont
  {Hendrych}, \citenamefont {Shi}, \citenamefont {Valencia},\ and\
  \citenamefont {Torres}}]{Hendrych09PRA}%
  \BibitemOpen
  \bibfield  {author} {\bibinfo {author} {\bibfnamefont {M.}~\bibnamefont
  {Hendrych}}, \bibinfo {author} {\bibfnamefont {X.}~\bibnamefont {Shi}},
  \bibinfo {author} {\bibfnamefont {A.}~\bibnamefont {Valencia}}, \ and\
  \bibinfo {author} {\bibfnamefont {J.~P.}\ \bibnamefont {Torres}},\
  }\href@noop {} {\bibfield  {journal} {\bibinfo  {journal} {Phys. Rev. A}\
  }\textbf {\bibinfo {volume} {79}},\ \bibinfo {pages} {023817} (\bibinfo
  {year} {2009})}\BibitemShut {NoStop}%
\bibitem [{\citenamefont {Hebling}\ \emph {et~al.}(2002)\citenamefont
  {Hebling}, \citenamefont {Almási}, \citenamefont {Kozma},\ and\
  \citenamefont {Kuhl}}]{Hebling02OE}%
  \BibitemOpen
  \bibfield  {author} {\bibinfo {author} {\bibfnamefont {J.}~\bibnamefont
  {Hebling}}, \bibinfo {author} {\bibfnamefont {G.}~\bibnamefont {Almási}},
  \bibinfo {author} {\bibfnamefont {I.~Z.}\ \bibnamefont {Kozma}}, \ and\
  \bibinfo {author} {\bibfnamefont {J.}~\bibnamefont {Kuhl}},\ }\href@noop {}
  {\bibfield  {journal} {\bibinfo  {journal} {Opt. Express}\ }\textbf {\bibinfo
  {volume} {10}},\ \bibinfo {pages} {1161} (\bibinfo {year}
  {2002})}\BibitemShut {NoStop}%
\bibitem [{\citenamefont {Hebling}\ \emph {et~al.}(2008)\citenamefont
  {Hebling}, \citenamefont {Yeh}, \citenamefont {Hoffmann}, \citenamefont
  {Bartal},\ and\ \citenamefont {Nelson}}]{Hebling08JOSAB}%
  \BibitemOpen
  \bibfield  {author} {\bibinfo {author} {\bibfnamefont {J.}~\bibnamefont
  {Hebling}}, \bibinfo {author} {\bibfnamefont {K.-L.}\ \bibnamefont {Yeh}},
  \bibinfo {author} {\bibfnamefont {M.~C.}\ \bibnamefont {Hoffmann}}, \bibinfo
  {author} {\bibfnamefont {B.}~\bibnamefont {Bartal}}, \ and\ \bibinfo {author}
  {\bibfnamefont {K.~A.}\ \bibnamefont {Nelson}},\ }\href@noop {} {\bibfield
  {journal} {\bibinfo  {journal} {J. Opt. Soc. Am. B}\ }\textbf {\bibinfo
  {volume} {25}},\ \bibinfo {pages} {B6} (\bibinfo {year} {2008})}\BibitemShut
  {NoStop}%
\bibitem [{\citenamefont {Wang}\ \emph {et~al.}(2020)\citenamefont {Wang},
  \citenamefont {T{\'o}th}, \citenamefont {Hebling},\ and\ \citenamefont
  {K{\"a}rtner}}]{Wang20LPR}%
  \BibitemOpen
  \bibfield  {author} {\bibinfo {author} {\bibfnamefont {L.}~\bibnamefont
  {Wang}}, \bibinfo {author} {\bibfnamefont {G.}~\bibnamefont {T{\'o}th}},
  \bibinfo {author} {\bibfnamefont {J.}~\bibnamefont {Hebling}}, \ and\
  \bibinfo {author} {\bibfnamefont {F.}~\bibnamefont {K{\"a}rtner}},\
  }\href@noop {} {\bibfield  {journal} {\bibinfo  {journal} {Laser Photon.
  Rev.}\ }\textbf {\bibinfo {volume} {14}},\ \bibinfo {pages} {2000021}
  (\bibinfo {year} {2020})}\BibitemShut {NoStop}%
\bibitem [{\citenamefont {Szatm{\'a}ri}\ \emph {et~al.}(1996)\citenamefont
  {Szatm{\'a}ri}, \citenamefont {Simon},\ and\ \citenamefont
  {Feuerhake}}]{Szatmari96OL}%
  \BibitemOpen
  \bibfield  {author} {\bibinfo {author} {\bibfnamefont {S.}~\bibnamefont
  {Szatm{\'a}ri}}, \bibinfo {author} {\bibfnamefont {P.}~\bibnamefont {Simon}},
  \ and\ \bibinfo {author} {\bibfnamefont {M.}~\bibnamefont {Feuerhake}},\
  }\href@noop {} {\bibfield  {journal} {\bibinfo  {journal} {Opt. Lett.}\
  }\textbf {\bibinfo {volume} {21}},\ \bibinfo {pages} {1156} (\bibinfo {year}
  {1996})}\BibitemShut {NoStop}%
\bibitem [{\citenamefont {Liu}\ and\ \citenamefont {Fan}(1998)}]{Liu98JMO}%
  \BibitemOpen
  \bibfield  {author} {\bibinfo {author} {\bibfnamefont {Z.}~\bibnamefont
  {Liu}}\ and\ \bibinfo {author} {\bibfnamefont {D.}~\bibnamefont {Fan}},\
  }\href@noop {} {\bibfield  {journal} {\bibinfo  {journal} {J. Mod. Opt.}\
  }\textbf {\bibinfo {volume} {45}},\ \bibinfo {pages} {17} (\bibinfo {year}
  {1998})}\BibitemShut {NoStop}%
\bibitem [{\citenamefont {Hu}\ and\ \citenamefont {Guo}(2002)}]{Hu02JOSAA}%
  \BibitemOpen
  \bibfield  {author} {\bibinfo {author} {\bibfnamefont {W.}~\bibnamefont
  {Hu}}\ and\ \bibinfo {author} {\bibfnamefont {H.}~\bibnamefont {Guo}},\
  }\href@noop {} {\bibfield  {journal} {\bibinfo  {journal} {J. Opt. Soc. Am.
  A}\ }\textbf {\bibinfo {volume} {19}},\ \bibinfo {pages} {49} (\bibinfo
  {year} {2002})}\BibitemShut {NoStop}%
\bibitem [{\citenamefont {L{\"u}}\ and\ \citenamefont {Liu}(2003)}]{Lu03JOSAA}%
  \BibitemOpen
  \bibfield  {author} {\bibinfo {author} {\bibfnamefont {B.}~\bibnamefont
  {L{\"u}}}\ and\ \bibinfo {author} {\bibfnamefont {Z.}~\bibnamefont {Liu}},\
  }\href@noop {} {\bibfield  {journal} {\bibinfo  {journal} {J. Opt. Soc. Am.
  A}\ }\textbf {\bibinfo {volume} {20}},\ \bibinfo {pages} {582} (\bibinfo
  {year} {2003})}\BibitemShut {NoStop}%
\bibitem [{\citenamefont {S{\~o}najalg}\ and\ \citenamefont
  {Saari}(1996)}]{Sonajalg96OL}%
  \BibitemOpen
  \bibfield  {author} {\bibinfo {author} {\bibfnamefont {H.}~\bibnamefont
  {S{\~o}najalg}}\ and\ \bibinfo {author} {\bibfnamefont {P.}~\bibnamefont
  {Saari}},\ }\href@noop {} {\bibfield  {journal} {\bibinfo  {journal} {Opt.
  Lett.}\ }\textbf {\bibinfo {volume} {21}},\ \bibinfo {pages} {1162} (\bibinfo
  {year} {1996})}\BibitemShut {NoStop}%
\bibitem [{\citenamefont {S{\~o}najalg}\ \emph {et~al.}(1997)\citenamefont
  {S{\~o}najalg}, \citenamefont {R{\"a}tsep},\ and\ \citenamefont
  {Saari}}]{Sonajalg97OL}%
  \BibitemOpen
  \bibfield  {author} {\bibinfo {author} {\bibfnamefont {H.}~\bibnamefont
  {S{\~o}najalg}}, \bibinfo {author} {\bibfnamefont {M.}~\bibnamefont
  {R{\"a}tsep}}, \ and\ \bibinfo {author} {\bibfnamefont {P.}~\bibnamefont
  {Saari}},\ }\href@noop {} {\bibfield  {journal} {\bibinfo  {journal} {Opt.
  Lett.}\ }\textbf {\bibinfo {volume} {22}},\ \bibinfo {pages} {310} (\bibinfo
  {year} {1997})}\BibitemShut {NoStop}%
\bibitem [{\citenamefont {Kondakci}\ and\ \citenamefont
  {Abouraddy}(2016)}]{Kondakci16OE}%
  \BibitemOpen
  \bibfield  {author} {\bibinfo {author} {\bibfnamefont {H.~E.}\ \bibnamefont
  {Kondakci}}\ and\ \bibinfo {author} {\bibfnamefont {A.~F.}\ \bibnamefont
  {Abouraddy}},\ }\href@noop {} {\bibfield  {journal} {\bibinfo  {journal}
  {Opt. Express}\ }\textbf {\bibinfo {volume} {24}},\ \bibinfo {pages} {28659}
  (\bibinfo {year} {2016})}\BibitemShut {NoStop}%
\bibitem [{\citenamefont {Parker}\ and\ \citenamefont
  {Alonso}(2016)}]{Parker16OE}%
  \BibitemOpen
  \bibfield  {author} {\bibinfo {author} {\bibfnamefont {K.~J.}\ \bibnamefont
  {Parker}}\ and\ \bibinfo {author} {\bibfnamefont {M.~A.}\ \bibnamefont
  {Alonso}},\ }\href@noop {} {\bibfield  {journal} {\bibinfo  {journal} {Opt.
  Express}\ }\textbf {\bibinfo {volume} {24}},\ \bibinfo {pages} {28669}
  (\bibinfo {year} {2016})}\BibitemShut {NoStop}%
\bibitem [{\citenamefont {Kondakci}\ and\ \citenamefont
  {Abouraddy}(2017)}]{Kondakci17NP}%
  \BibitemOpen
  \bibfield  {author} {\bibinfo {author} {\bibfnamefont {H.~E.}\ \bibnamefont
  {Kondakci}}\ and\ \bibinfo {author} {\bibfnamefont {A.~F.}\ \bibnamefont
  {Abouraddy}},\ }\href@noop {} {\bibfield  {journal} {\bibinfo  {journal}
  {Nat. Photon.}\ }\textbf {\bibinfo {volume} {11}},\ \bibinfo {pages} {733}
  (\bibinfo {year} {2017})}\BibitemShut {NoStop}%
\bibitem [{\citenamefont {Yessenov}\ \emph
  {et~al.}(2019{\natexlab{a}})\citenamefont {Yessenov}, \citenamefont
  {Bhaduri}, \citenamefont {Kondakci},\ and\ \citenamefont
  {Abouraddy}}]{Yessenov19OPN}%
  \BibitemOpen
  \bibfield  {author} {\bibinfo {author} {\bibfnamefont {M.}~\bibnamefont
  {Yessenov}}, \bibinfo {author} {\bibfnamefont {B.}~\bibnamefont {Bhaduri}},
  \bibinfo {author} {\bibfnamefont {H.~E.}\ \bibnamefont {Kondakci}}, \ and\
  \bibinfo {author} {\bibfnamefont {A.~F.}\ \bibnamefont {Abouraddy}},\
  }\href@noop {} {\bibfield  {journal} {\bibinfo  {journal} {Opt. Photon.
  News}\ }\textbf {\bibinfo {volume} {30}},\ \bibinfo {pages} {34} (\bibinfo
  {year} {2019}{\natexlab{a}})}\BibitemShut {NoStop}%
\bibitem [{\citenamefont {Donnelly}\ and\ \citenamefont
  {Ziolkowski}(1993)}]{Donnelly93ProcRSLA}%
  \BibitemOpen
  \bibfield  {author} {\bibinfo {author} {\bibfnamefont {R.}~\bibnamefont
  {Donnelly}}\ and\ \bibinfo {author} {\bibfnamefont {R.~W.}\ \bibnamefont
  {Ziolkowski}},\ }\href@noop {} {\bibfield  {journal} {\bibinfo  {journal}
  {Proc. R. Soc. Lond. A}\ }\textbf {\bibinfo {volume} {440}},\ \bibinfo
  {pages} {541} (\bibinfo {year} {1993})}\BibitemShut {NoStop}%
\bibitem [{\citenamefont {Saari}\ and\ \citenamefont
  {Reivelt}(2004)}]{Saari04PRE}%
  \BibitemOpen
  \bibfield  {author} {\bibinfo {author} {\bibfnamefont {P.}~\bibnamefont
  {Saari}}\ and\ \bibinfo {author} {\bibfnamefont {K.}~\bibnamefont
  {Reivelt}},\ }\href@noop {} {\bibfield  {journal} {\bibinfo  {journal} {Phys.
  Rev. E}\ }\textbf {\bibinfo {volume} {69}},\ \bibinfo {pages} {036612}
  (\bibinfo {year} {2004})}\BibitemShut {NoStop}%
\bibitem [{\citenamefont {Longhi}(2004{\natexlab{a}})}]{Longhi04OE}%
  \BibitemOpen
  \bibfield  {author} {\bibinfo {author} {\bibfnamefont {S.}~\bibnamefont
  {Longhi}},\ }\href@noop {} {\bibfield  {journal} {\bibinfo  {journal} {Opt.
  Express}\ }\textbf {\bibinfo {volume} {12}},\ \bibinfo {pages} {935}
  (\bibinfo {year} {2004}{\natexlab{a}})}\BibitemShut {NoStop}%
\bibitem [{\citenamefont {Valtna}\ \emph {et~al.}(2007)\citenamefont {Valtna},
  \citenamefont {Reivelt},\ and\ \citenamefont {Saari}}]{Valtna07OC}%
  \BibitemOpen
  \bibfield  {author} {\bibinfo {author} {\bibfnamefont {H.}~\bibnamefont
  {Valtna}}, \bibinfo {author} {\bibfnamefont {K.}~\bibnamefont {Reivelt}}, \
  and\ \bibinfo {author} {\bibfnamefont {P.}~\bibnamefont {Saari}},\
  }\href@noop {} {\bibfield  {journal} {\bibinfo  {journal} {Opt. Commun.}\
  }\textbf {\bibinfo {volume} {278}},\ \bibinfo {pages} {1} (\bibinfo {year}
  {2007})}\BibitemShut {NoStop}%
\bibitem [{\citenamefont {Wong}\ and\ \citenamefont
  {Kaminer}(2017{\natexlab{a}})}]{Wong17ACSP1}%
  \BibitemOpen
  \bibfield  {author} {\bibinfo {author} {\bibfnamefont {L.~J.}\ \bibnamefont
  {Wong}}\ and\ \bibinfo {author} {\bibfnamefont {I.}~\bibnamefont {Kaminer}},\
  }\href@noop {} {\bibfield  {journal} {\bibinfo  {journal} {ACS Photon.}\
  }\textbf {\bibinfo {volume} {4}},\ \bibinfo {pages} {1131} (\bibinfo {year}
  {2017}{\natexlab{a}})}\BibitemShut {NoStop}%
\bibitem [{\citenamefont {Wong}\ and\ \citenamefont
  {Kaminer}(2017{\natexlab{b}})}]{Wong17ACSP2}%
  \BibitemOpen
  \bibfield  {author} {\bibinfo {author} {\bibfnamefont {L.~J.}\ \bibnamefont
  {Wong}}\ and\ \bibinfo {author} {\bibfnamefont {I.}~\bibnamefont {Kaminer}},\
  }\href@noop {} {\bibfield  {journal} {\bibinfo  {journal} {ACS Photon.}\
  }\textbf {\bibinfo {volume} {4}},\ \bibinfo {pages} {2257} (\bibinfo {year}
  {2017}{\natexlab{b}})}\BibitemShut {NoStop}%
\bibitem [{\citenamefont {Porras}(2017)}]{Porras17OL}%
  \BibitemOpen
  \bibfield  {author} {\bibinfo {author} {\bibfnamefont {M.~A.}\ \bibnamefont
  {Porras}},\ }\href@noop {} {\bibfield  {journal} {\bibinfo  {journal} {Opt.
  Lett.}\ }\textbf {\bibinfo {volume} {42}},\ \bibinfo {pages} {4679} (\bibinfo
  {year} {2017})}\BibitemShut {NoStop}%
\bibitem [{\citenamefont {Efremidis}(2017)}]{Efremidis17OL}%
  \BibitemOpen
  \bibfield  {author} {\bibinfo {author} {\bibfnamefont {N.~K.}\ \bibnamefont
  {Efremidis}},\ }\href@noop {} {\bibfield  {journal} {\bibinfo  {journal}
  {Opt. Lett.}\ }\textbf {\bibinfo {volume} {42}},\ \bibinfo {pages} {5038}
  (\bibinfo {year} {2017})}\BibitemShut {NoStop}%
\bibitem [{\citenamefont {Wong}\ \emph {et~al.}(2020)\citenamefont {Wong},
  \citenamefont {Christodoulides},\ and\ \citenamefont {Kaminer}}]{Wong20AS}%
  \BibitemOpen
  \bibfield  {author} {\bibinfo {author} {\bibfnamefont {L.~J.}\ \bibnamefont
  {Wong}}, \bibinfo {author} {\bibfnamefont {D.~N.}\ \bibnamefont
  {Christodoulides}}, \ and\ \bibinfo {author} {\bibfnamefont {I.}~\bibnamefont
  {Kaminer}},\ }\href@noop {} {\bibfield  {journal} {\bibinfo  {journal} {Adv.
  Sci.}\ }\textbf {\bibinfo {volume} {7}},\ \bibinfo {pages} {1903377}
  (\bibinfo {year} {2020})}\BibitemShut {NoStop}%
\bibitem [{\citenamefont {Hall}\ \emph {et~al.}(2021)\citenamefont {Hall},
  \citenamefont {Yessenov},\ and\ \citenamefont {Abouraddy}}]{Hall21arxiv}%
  \BibitemOpen
  \bibfield  {author} {\bibinfo {author} {\bibfnamefont {L.~A.}\ \bibnamefont
  {Hall}}, \bibinfo {author} {\bibfnamefont {M.}~\bibnamefont {Yessenov}}, \
  and\ \bibinfo {author} {\bibfnamefont {A.~F.}\ \bibnamefont {Abouraddy}},\
  }\href@noop {} {\bibfield  {journal} {\bibinfo  {journal} {arXiv:2101.07317}\
  } (\bibinfo {year} {2021})}\BibitemShut {NoStop}%
\bibitem [{\citenamefont {Bhaduri}\ \emph {et~al.}(2018)\citenamefont
  {Bhaduri}, \citenamefont {Yessenov},\ and\ \citenamefont
  {Abouraddy}}]{Bhaduri18OE}%
  \BibitemOpen
  \bibfield  {author} {\bibinfo {author} {\bibfnamefont {B.}~\bibnamefont
  {Bhaduri}}, \bibinfo {author} {\bibfnamefont {M.}~\bibnamefont {Yessenov}}, \
  and\ \bibinfo {author} {\bibfnamefont {A.~F.}\ \bibnamefont {Abouraddy}},\
  }\href@noop {} {\bibfield  {journal} {\bibinfo  {journal} {Opt. Express}\
  }\textbf {\bibinfo {volume} {26}},\ \bibinfo {pages} {20111} (\bibinfo {year}
  {2018})}\BibitemShut {NoStop}%
\bibitem [{\citenamefont {Bhaduri}\ \emph
  {et~al.}(2019{\natexlab{a}})\citenamefont {Bhaduri}, \citenamefont
  {Yessenov}, \citenamefont {Reyes}, \citenamefont {Pena}, \citenamefont
  {Meem}, \citenamefont {Fairchild}, \citenamefont {Menon}, \citenamefont
  {Richardson},\ and\ \citenamefont {Abouraddy}}]{Bhaduri19OL}%
  \BibitemOpen
  \bibfield  {author} {\bibinfo {author} {\bibfnamefont {B.}~\bibnamefont
  {Bhaduri}}, \bibinfo {author} {\bibfnamefont {M.}~\bibnamefont {Yessenov}},
  \bibinfo {author} {\bibfnamefont {D.}~\bibnamefont {Reyes}}, \bibinfo
  {author} {\bibfnamefont {J.}~\bibnamefont {Pena}}, \bibinfo {author}
  {\bibfnamefont {M.}~\bibnamefont {Meem}}, \bibinfo {author} {\bibfnamefont
  {S.~R.}\ \bibnamefont {Fairchild}}, \bibinfo {author} {\bibfnamefont
  {R.}~\bibnamefont {Menon}}, \bibinfo {author} {\bibfnamefont {M.~C.}\
  \bibnamefont {Richardson}}, \ and\ \bibinfo {author} {\bibfnamefont {A.~F.}\
  \bibnamefont {Abouraddy}},\ }\href@noop {} {\bibfield  {journal} {\bibinfo
  {journal} {Opt. Lett.}\ }\textbf {\bibinfo {volume} {44}},\ \bibinfo {pages}
  {2073} (\bibinfo {year} {2019}{\natexlab{a}})}\BibitemShut {NoStop}%
\bibitem [{\citenamefont {Schepler}\ \emph {et~al.}(2020)\citenamefont
  {Schepler}, \citenamefont {Yessenov}, \citenamefont {Zhiyenbayev},\ and\
  \citenamefont {Abouraddy}}]{Schepler20ACSP}%
  \BibitemOpen
  \bibfield  {author} {\bibinfo {author} {\bibfnamefont {K.~L.}\ \bibnamefont
  {Schepler}}, \bibinfo {author} {\bibfnamefont {M.}~\bibnamefont {Yessenov}},
  \bibinfo {author} {\bibfnamefont {Y.}~\bibnamefont {Zhiyenbayev}}, \ and\
  \bibinfo {author} {\bibfnamefont {A.~F.}\ \bibnamefont {Abouraddy}},\
  }\href@noop {} {\bibfield  {journal} {\bibinfo  {journal} {ACS Photon.}\
  }\textbf {\bibinfo {volume} {7}},\ \bibinfo {pages} {2966} (\bibinfo {year}
  {2020})}\BibitemShut {NoStop}%
\bibitem [{\citenamefont {Yessenov}\ \emph
  {et~al.}(2020{\natexlab{a}})\citenamefont {Yessenov}, \citenamefont
  {Bhaduri}, \citenamefont {Delfyett},\ and\ \citenamefont
  {Abouraddy}}]{Yessenov20NC}%
  \BibitemOpen
  \bibfield  {author} {\bibinfo {author} {\bibfnamefont {M.}~\bibnamefont
  {Yessenov}}, \bibinfo {author} {\bibfnamefont {B.}~\bibnamefont {Bhaduri}},
  \bibinfo {author} {\bibfnamefont {P.~J.}\ \bibnamefont {Delfyett}}, \ and\
  \bibinfo {author} {\bibfnamefont {A.~F.}\ \bibnamefont {Abouraddy}},\
  }\href@noop {} {\bibfield  {journal} {\bibinfo  {journal} {Nat. Commun.}\
  }\textbf {\bibinfo {volume} {11}},\ \bibinfo {pages} {5782} (\bibinfo {year}
  {2020}{\natexlab{a}})}\BibitemShut {NoStop}%
\bibitem [{\citenamefont {Shiri}\ \emph {et~al.}(2020)\citenamefont {Shiri},
  \citenamefont {Yessenov}, \citenamefont {Webster}, \citenamefont {Schepler},\
  and\ \citenamefont {Abouraddy}}]{Shiri20NC}%
  \BibitemOpen
  \bibfield  {author} {\bibinfo {author} {\bibfnamefont {A.}~\bibnamefont
  {Shiri}}, \bibinfo {author} {\bibfnamefont {M.}~\bibnamefont {Yessenov}},
  \bibinfo {author} {\bibfnamefont {S.}~\bibnamefont {Webster}}, \bibinfo
  {author} {\bibfnamefont {K.~L.}\ \bibnamefont {Schepler}}, \ and\ \bibinfo
  {author} {\bibfnamefont {A.~F.}\ \bibnamefont {Abouraddy}},\ }\href@noop {}
  {\bibfield  {journal} {\bibinfo  {journal} {Nat. Commun.}\ }\textbf {\bibinfo
  {volume} {11}},\ \bibinfo {pages} {6273} (\bibinfo {year}
  {2020})}\BibitemShut {NoStop}%
\bibitem [{\citenamefont {Salo}\ and\ \citenamefont
  {Salomaa}(2001)}]{Salo01JOA}%
  \BibitemOpen
  \bibfield  {author} {\bibinfo {author} {\bibfnamefont {J.}~\bibnamefont
  {Salo}}\ and\ \bibinfo {author} {\bibfnamefont {M.~M.}\ \bibnamefont
  {Salomaa}},\ }\href@noop {} {\bibfield  {journal} {\bibinfo  {journal} {J.
  Opt. A}\ }\textbf {\bibinfo {volume} {3}},\ \bibinfo {pages} {366} (\bibinfo
  {year} {2001})}\BibitemShut {NoStop}%
\bibitem [{\citenamefont {Kondakci}\ and\ \citenamefont
  {Abouraddy}(2019)}]{Kondakci19NC}%
  \BibitemOpen
  \bibfield  {author} {\bibinfo {author} {\bibfnamefont {H.~E.}\ \bibnamefont
  {Kondakci}}\ and\ \bibinfo {author} {\bibfnamefont {A.~F.}\ \bibnamefont
  {Abouraddy}},\ }\href@noop {} {\bibfield  {journal} {\bibinfo  {journal}
  {Nat. Commun.}\ }\textbf {\bibinfo {volume} {10}},\ \bibinfo {pages} {929}
  (\bibinfo {year} {2019})}\BibitemShut {NoStop}%
\bibitem [{\citenamefont {Bhaduri}\ \emph
  {et~al.}(2019{\natexlab{b}})\citenamefont {Bhaduri}, \citenamefont
  {Yessenov},\ and\ \citenamefont {Abouraddy}}]{Bhaduri19Optica}%
  \BibitemOpen
  \bibfield  {author} {\bibinfo {author} {\bibfnamefont {B.}~\bibnamefont
  {Bhaduri}}, \bibinfo {author} {\bibfnamefont {M.}~\bibnamefont {Yessenov}}, \
  and\ \bibinfo {author} {\bibfnamefont {A.~F.}\ \bibnamefont {Abouraddy}},\
  }\href@noop {} {\bibfield  {journal} {\bibinfo  {journal} {Optica}\ }\textbf
  {\bibinfo {volume} {6}},\ \bibinfo {pages} {139} (\bibinfo {year}
  {2019}{\natexlab{b}})}\BibitemShut {NoStop}%
\bibitem [{\citenamefont {Yessenov}\ \emph
  {et~al.}(2019{\natexlab{b}})\citenamefont {Yessenov}, \citenamefont
  {Bhaduri}, \citenamefont {Mach}, \citenamefont {Mardani}, \citenamefont
  {Kondakci}, \citenamefont {Alonso}, \citenamefont {Atia},\ and\ \citenamefont
  {Abouraddy}}]{Yessenov19OE}%
  \BibitemOpen
  \bibfield  {author} {\bibinfo {author} {\bibfnamefont {M.}~\bibnamefont
  {Yessenov}}, \bibinfo {author} {\bibfnamefont {B.}~\bibnamefont {Bhaduri}},
  \bibinfo {author} {\bibfnamefont {L.}~\bibnamefont {Mach}}, \bibinfo {author}
  {\bibfnamefont {D.}~\bibnamefont {Mardani}}, \bibinfo {author} {\bibfnamefont
  {H.~E.}\ \bibnamefont {Kondakci}}, \bibinfo {author} {\bibfnamefont {M.~A.}\
  \bibnamefont {Alonso}}, \bibinfo {author} {\bibfnamefont {G.~A.}\
  \bibnamefont {Atia}}, \ and\ \bibinfo {author} {\bibfnamefont {A.~F.}\
  \bibnamefont {Abouraddy}},\ }\href@noop {} {\bibfield  {journal} {\bibinfo
  {journal} {Opt. Express}\ }\textbf {\bibinfo {volume} {27}},\ \bibinfo
  {pages} {12443} (\bibinfo {year} {2019}{\natexlab{b}})}\BibitemShut {NoStop}%
\bibitem [{\citenamefont {Bhaduri}\ \emph {et~al.}(2020)\citenamefont
  {Bhaduri}, \citenamefont {Yessenov},\ and\ \citenamefont
  {Abouraddy}}]{Bhaduri20NP}%
  \BibitemOpen
  \bibfield  {author} {\bibinfo {author} {\bibfnamefont {B.}~\bibnamefont
  {Bhaduri}}, \bibinfo {author} {\bibfnamefont {M.}~\bibnamefont {Yessenov}}, \
  and\ \bibinfo {author} {\bibfnamefont {A.~F.}\ \bibnamefont {Abouraddy}},\
  }\href@noop {} {\bibfield  {journal} {\bibinfo  {journal} {Nat. Photon.}\
  }\textbf {\bibinfo {volume} {14}},\ \bibinfo {pages} {416} (\bibinfo {year}
  {2020})}\BibitemShut {NoStop}%
\bibitem [{\citenamefont {Kondakci}\ \emph {et~al.}(2018)\citenamefont
  {Kondakci}, \citenamefont {Yessenov}, \citenamefont {Meem}, \citenamefont
  {Reyes}, \citenamefont {Thul}, \citenamefont {Fairchild}, \citenamefont
  {Richardson}, \citenamefont {Menon},\ and\ \citenamefont
  {Abouraddy}}]{Kondakci18OE}%
  \BibitemOpen
  \bibfield  {author} {\bibinfo {author} {\bibfnamefont {H.~E.}\ \bibnamefont
  {Kondakci}}, \bibinfo {author} {\bibfnamefont {M.}~\bibnamefont {Yessenov}},
  \bibinfo {author} {\bibfnamefont {M.}~\bibnamefont {Meem}}, \bibinfo {author}
  {\bibfnamefont {D.}~\bibnamefont {Reyes}}, \bibinfo {author} {\bibfnamefont
  {D.}~\bibnamefont {Thul}}, \bibinfo {author} {\bibfnamefont {S.~R.}\
  \bibnamefont {Fairchild}}, \bibinfo {author} {\bibfnamefont {M.}~\bibnamefont
  {Richardson}}, \bibinfo {author} {\bibfnamefont {R.}~\bibnamefont {Menon}}, \
  and\ \bibinfo {author} {\bibfnamefont {A.~F.}\ \bibnamefont {Abouraddy}},\
  }\href@noop {} {\bibfield  {journal} {\bibinfo  {journal} {Opt. Express}\
  }\textbf {\bibinfo {volume} {26}},\ \bibinfo {pages} {13628} (\bibinfo {year}
  {2018})}\BibitemShut {NoStop}%
\bibitem [{\citenamefont {Yessenov}\ \emph
  {et~al.}(2020{\natexlab{b}})\citenamefont {Yessenov}, \citenamefont {Ru},
  \citenamefont {Schepler}, \citenamefont {Meem}, \citenamefont {Menon},
  \citenamefont {Vodopyanov},\ and\ \citenamefont
  {Abouraddy}}]{Yessenov20OSAC}%
  \BibitemOpen
  \bibfield  {author} {\bibinfo {author} {\bibfnamefont {M.}~\bibnamefont
  {Yessenov}}, \bibinfo {author} {\bibfnamefont {Q.}~\bibnamefont {Ru}},
  \bibinfo {author} {\bibfnamefont {K.~L.}\ \bibnamefont {Schepler}}, \bibinfo
  {author} {\bibfnamefont {M.}~\bibnamefont {Meem}}, \bibinfo {author}
  {\bibfnamefont {R.}~\bibnamefont {Menon}}, \bibinfo {author} {\bibfnamefont
  {K.~L.}\ \bibnamefont {Vodopyanov}}, \ and\ \bibinfo {author} {\bibfnamefont
  {A.~F.}\ \bibnamefont {Abouraddy}},\ }\href@noop {} {\bibfield  {journal}
  {\bibinfo  {journal} {OSA Continuum}\ }\textbf {\bibinfo {volume} {3}},\
  \bibinfo {pages} {420} (\bibinfo {year} {2020}{\natexlab{b}})}\BibitemShut
  {NoStop}%
\bibitem [{\citenamefont {Reivelt}\ and\ \citenamefont
  {Saari}(2003)}]{Reivelt03arxiv}%
  \BibitemOpen
  \bibfield  {author} {\bibinfo {author} {\bibfnamefont {K.}~\bibnamefont
  {Reivelt}}\ and\ \bibinfo {author} {\bibfnamefont {P.}~\bibnamefont
  {Saari}},\ }\href@noop {} {\bibfield  {journal} {\bibinfo  {journal}
  {arxiv:physics/0309079}\ } (\bibinfo {year} {2003})}\BibitemShut {NoStop}%
\bibitem [{\citenamefont {Kiselev}(2007)}]{Kiselev07OS}%
  \BibitemOpen
  \bibfield  {author} {\bibinfo {author} {\bibfnamefont {A.~P.}\ \bibnamefont
  {Kiselev}},\ }\href@noop {} {\bibfield  {journal} {\bibinfo  {journal} {Opt.
  Spectrosc.}\ }\textbf {\bibinfo {volume} {102}},\ \bibinfo {pages} {603}
  (\bibinfo {year} {2007})}\BibitemShut {NoStop}%
\bibitem [{\citenamefont {Turunen}\ and\ \citenamefont
  {Friberg}(2010)}]{Turunen10PO}%
  \BibitemOpen
  \bibfield  {author} {\bibinfo {author} {\bibfnamefont {J.}~\bibnamefont
  {Turunen}}\ and\ \bibinfo {author} {\bibfnamefont {A.~T.}\ \bibnamefont
  {Friberg}},\ }\href@noop {} {\bibfield  {journal} {\bibinfo  {journal} {Prog.
  Opt.}\ }\textbf {\bibinfo {volume} {54}},\ \bibinfo {pages} {1} (\bibinfo
  {year} {2010})}\BibitemShut {NoStop}%
\bibitem [{\citenamefont {Hern\'andez-Figueroa}\ \emph
  {et~al.}(2014)\citenamefont {Hern\'andez-Figueroa}, \citenamefont {Recami},\
  and\ \citenamefont {Zamboni-Rached}}]{FigueroaBook14}%
  \BibitemOpen
  \bibinfo {editor} {\bibfnamefont {H.~E.}\ \bibnamefont
  {Hern\'andez-Figueroa}}, \bibinfo {editor} {\bibfnamefont {E.}~\bibnamefont
  {Recami}}, \ and\ \bibinfo {editor} {\bibfnamefont {M.}~\bibnamefont
  {Zamboni-Rached}},\ eds.,\ \href@noop {} {\emph {\bibinfo {title}
  {Non-diffracting Waves}}}\ (\bibinfo  {publisher} {Wiley-VCH},\ \bibinfo
  {year} {2014})\BibitemShut {NoStop}%
\bibitem [{\citenamefont {Yessenov}\ \emph
  {et~al.}(2019{\natexlab{c}})\citenamefont {Yessenov}, \citenamefont
  {Bhaduri}, \citenamefont {Kondakci},\ and\ \citenamefont
  {Abouraddy}}]{Yessenov19PRA}%
  \BibitemOpen
  \bibfield  {author} {\bibinfo {author} {\bibfnamefont {M.}~\bibnamefont
  {Yessenov}}, \bibinfo {author} {\bibfnamefont {B.}~\bibnamefont {Bhaduri}},
  \bibinfo {author} {\bibfnamefont {H.~E.}\ \bibnamefont {Kondakci}}, \ and\
  \bibinfo {author} {\bibfnamefont {A.~F.}\ \bibnamefont {Abouraddy}},\
  }\href@noop {} {\bibfield  {journal} {\bibinfo  {journal} {Phys. Rev. A}\
  }\textbf {\bibinfo {volume} {99}},\ \bibinfo {pages} {023856} (\bibinfo
  {year} {2019}{\natexlab{c}})}\BibitemShut {NoStop}%
\bibitem [{\citenamefont {Kondakci}\ \emph {et~al.}(2019)\citenamefont
  {Kondakci}, \citenamefont {Nye}, \citenamefont {Christodoulides},\ and\
  \citenamefont {Abouraddy}}]{Kondakci19ACSP}%
  \BibitemOpen
  \bibfield  {author} {\bibinfo {author} {\bibfnamefont {H.~E.}\ \bibnamefont
  {Kondakci}}, \bibinfo {author} {\bibfnamefont {N.~S.}\ \bibnamefont {Nye}},
  \bibinfo {author} {\bibfnamefont {D.~N.}\ \bibnamefont {Christodoulides}}, \
  and\ \bibinfo {author} {\bibfnamefont {A.~F.}\ \bibnamefont {Abouraddy}},\
  }\href@noop {} {\bibfield  {journal} {\bibinfo  {journal} {ACS Photon.}\
  }\textbf {\bibinfo {volume} {6}},\ \bibinfo {pages} {475} (\bibinfo {year}
  {2019})}\BibitemShut {NoStop}%
\bibitem [{\citenamefont {Yessenov}\ \emph
  {et~al.}(2019{\natexlab{d}})\citenamefont {Yessenov}, \citenamefont
  {Bhaduri}, \citenamefont {Kondakci}, \citenamefont {Meem}, \citenamefont
  {Menon},\ and\ \citenamefont {Abouraddy}}]{Yessenov19Optica}%
  \BibitemOpen
  \bibfield  {author} {\bibinfo {author} {\bibfnamefont {M.}~\bibnamefont
  {Yessenov}}, \bibinfo {author} {\bibfnamefont {B.}~\bibnamefont {Bhaduri}},
  \bibinfo {author} {\bibfnamefont {H.~E.}\ \bibnamefont {Kondakci}}, \bibinfo
  {author} {\bibfnamefont {M.}~\bibnamefont {Meem}}, \bibinfo {author}
  {\bibfnamefont {R.}~\bibnamefont {Menon}}, \ and\ \bibinfo {author}
  {\bibfnamefont {A.~F.}\ \bibnamefont {Abouraddy}},\ }\href@noop {} {\bibfield
   {journal} {\bibinfo  {journal} {Optica}\ }\textbf {\bibinfo {volume} {6}},\
  \bibinfo {pages} {522} (\bibinfo {year} {2019}{\natexlab{d}})}\BibitemShut
  {NoStop}%
\bibitem [{\citenamefont {Yessenov}\ and\ \citenamefont
  {Abouraddy}(2019)}]{Yessenov19OL}%
  \BibitemOpen
  \bibfield  {author} {\bibinfo {author} {\bibfnamefont {M.}~\bibnamefont
  {Yessenov}}\ and\ \bibinfo {author} {\bibfnamefont {A.~F.}\ \bibnamefont
  {Abouraddy}},\ }\href@noop {} {\bibfield  {journal} {\bibinfo  {journal}
  {Opt. Lett.}\ }\textbf {\bibinfo {volume} {44}},\ \bibinfo {pages} {5125}
  (\bibinfo {year} {2019})}\BibitemShut {NoStop}%
\bibitem [{\citenamefont {Saari}\ and\ \citenamefont
  {Reivelt}(1997)}]{Saari97PRL}%
  \BibitemOpen
  \bibfield  {author} {\bibinfo {author} {\bibfnamefont {P.}~\bibnamefont
  {Saari}}\ and\ \bibinfo {author} {\bibfnamefont {K.}~\bibnamefont
  {Reivelt}},\ }\href@noop {} {\bibfield  {journal} {\bibinfo  {journal} {Phys.
  Rev. Lett.}\ }\textbf {\bibinfo {volume} {79}},\ \bibinfo {pages} {4135}
  (\bibinfo {year} {1997})}\BibitemShut {NoStop}%
\bibitem [{\citenamefont {Brittingham}(1983)}]{Brittingham83JAP}%
  \BibitemOpen
  \bibfield  {author} {\bibinfo {author} {\bibfnamefont {J.~N.}\ \bibnamefont
  {Brittingham}},\ }\href@noop {} {\bibfield  {journal} {\bibinfo  {journal}
  {J. Appl. Phys.}\ }\textbf {\bibinfo {volume} {54}},\ \bibinfo {pages} {1179}
  (\bibinfo {year} {1983})}\BibitemShut {NoStop}%
\bibitem [{\citenamefont {Reivelt}\ and\ \citenamefont
  {Saari}(2000)}]{Reivelt00JOSAA}%
  \BibitemOpen
  \bibfield  {author} {\bibinfo {author} {\bibfnamefont {K.}~\bibnamefont
  {Reivelt}}\ and\ \bibinfo {author} {\bibfnamefont {P.}~\bibnamefont
  {Saari}},\ }\href@noop {} {\bibfield  {journal} {\bibinfo  {journal} {J. Opt.
  Soc. Am. A}\ }\textbf {\bibinfo {volume} {17}},\ \bibinfo {pages} {1785}
  (\bibinfo {year} {2000})}\BibitemShut {NoStop}%
\bibitem [{\citenamefont {Reivelt}\ and\ \citenamefont
  {Saari}(2002)}]{Reivelt02PRE}%
  \BibitemOpen
  \bibfield  {author} {\bibinfo {author} {\bibfnamefont {K.}~\bibnamefont
  {Reivelt}}\ and\ \bibinfo {author} {\bibfnamefont {P.}~\bibnamefont
  {Saari}},\ }\href@noop {} {\bibfield  {journal} {\bibinfo  {journal} {Phys.
  Rev. E}\ }\textbf {\bibinfo {volume} {66}},\ \bibinfo {pages} {056611}
  (\bibinfo {year} {2002})}\BibitemShut {NoStop}%
\bibitem [{\citenamefont {Motz}\ \emph {et~al.}(2020)\citenamefont {Motz},
  \citenamefont {Yessenov},\ and\ \citenamefont {Abouraddy}}]{Motz20arxiv}%
  \BibitemOpen
  \bibfield  {author} {\bibinfo {author} {\bibfnamefont {A.~M.~A.}\
  \bibnamefont {Motz}}, \bibinfo {author} {\bibfnamefont {M.}~\bibnamefont
  {Yessenov}}, \ and\ \bibinfo {author} {\bibfnamefont {A.~F.}\ \bibnamefont
  {Abouraddy}},\ }\href@noop {} {\bibfield  {journal} {\bibinfo  {journal}
  {arXiv:2010.10719}\ } (\bibinfo {year} {2020})}\BibitemShut {NoStop}%
\bibitem [{\citenamefont {Yessenov}\ and\ \citenamefont
  {Abouraddy}(2020)}]{Yessenov20PRL2}%
  \BibitemOpen
  \bibfield  {author} {\bibinfo {author} {\bibfnamefont {M.}~\bibnamefont
  {Yessenov}}\ and\ \bibinfo {author} {\bibfnamefont {A.~F.}\ \bibnamefont
  {Abouraddy}},\ }\href@noop {} {\bibfield  {journal} {\bibinfo  {journal}
  {Phys. Rev. Lett.}\ }\textbf {\bibinfo {volume} {125}},\ \bibinfo {pages}
  {233901} (\bibinfo {year} {2020})}\BibitemShut {NoStop}%
\bibitem [{\citenamefont {Longhi}(2004{\natexlab{b}})}]{Longhi04OL}%
  \BibitemOpen
  \bibfield  {author} {\bibinfo {author} {\bibfnamefont {S.}~\bibnamefont
  {Longhi}},\ }\href@noop {} {\bibfield  {journal} {\bibinfo  {journal} {Opt.
  Lett.}\ }\textbf {\bibinfo {volume} {29}},\ \bibinfo {pages} {147} (\bibinfo
  {year} {2004}{\natexlab{b}})}\BibitemShut {NoStop}%
\end{thebibliography}%


\end{document}